\documentclass[trackchanges]{aastex631}
\usepackage{bm}
\usepackage{amsmath}
\usepackage{multirow}

\renewcommand{\vec}[1]{\boldsymbol{#1}}
\usepackage{lipsum}

\begin{document}

\title{Wave-particle equilibria with heavy ions in weakly collisional space plasmas}

\correspondingauthor{Nicol\'as Villarroel-Sep\'ulveda}
\email{nicolas.villarroel@ug.uchile.cl}

\author[0000-0003-4887-9512]{Nicol\'as Villarroel-Sep\'ulveda}
\affiliation{Departamento de F\'isica, Facultad de Ciencias, Universidad de Chile,
	Las Palmeras 3425, \~Nu\~noa 7800003, Santiago, Chile}
\affiliation{Mullard Space Science Laboratory, University College London, Dorking RH5 6NT, United Kingdom}

\author[0000-0002-0497-1096]{Daniel Verscharen}
\affiliation{Mullard Space Science Laboratory, University College London, Dorking RH5 6NT, United Kingdom}

\author[0000-0002-9161-0888]{Pablo S. Moya}
\affiliation{Departamento de F\'isica, Facultad de Ciencias, Universidad de Chile, Las Palmeras 3425, \~Nu\~noa 7800003, Santiago, Chile}

\author[0000-0003-3223-1498]{Rodrigo A. L\'opez}
\affiliation{Research Center in the Intersection of Plasma Physics, Matter, and Complexity (P²mc), Comisión Chilena de Energ\'ia Nuclear, Casilla 188-D, Santiago, Chile}
\affiliation{Departamento de Ciencias F\'{\i}sicas, Facultad de Ciencias Exactas, Universidad Andres Bello, Sazi\'e 2212, Santiago 8370136, Chile}

\author[0000-0001-6038-1923]{Kristopher G. Klein}
\affiliation{Department of Planetary Sciences, Lunar and Planetary Laboratory, University of Arizona, 1629 E University Blvd, Tucson, AZ 85721, USA}

\title{Wave-particle equilibria with heavy ions in weakly collisional space plasmas}

\begin{abstract}
Space plasmas are weakly collisional since characteristic time scales related to Coulomb collisions are much larger than those of Larmor gyration or wave--particle interactions. Thus, wave activity is likely to drive some of the non-thermal features that are observed in space plasma velocity distributions, such as temperature anisotropy, beams, and skewness. Therefore, we study how wave--particle interactions shape the velocity distribution functions of minor ions, and how these ions and their statistical properties modify the dispersion relation of electromagnetic waves.
To achieve this, we derive the motion of heavy ions in electromagnetic waves using the Boris algorithm. We take the waves to be solutions of the fully kinetic dispersion relation of
electromagnetic waves in two-ion component plasmas with parameters representative of the solar wind. We use the Arbitrary Linear Plasma Solver (ALPS) code to derive the linear Vlasov--Maxwell dispersion relation based on the actual distribution of the ions. The test-particles are initially in thermal equilibrium, and their distribution evolves due to interactions with the waves. By solving the dispersion relation using the evolved distributions, we show that the system evolves into a steady wave--particle equilibrium, which is characterized by a minimization of the interaction and energy transfer between wave and particles.
\end{abstract}
\section{Introduction} \label{sec:intro}
Protons are the most abundant ion species in most space plasmas, but heavier ions in multiple ionization states are also present in smaller concentrations \citep{Neugebauer_1966, Bame_1968, Boschler2007}. In particular, the most common heavy ions in the solar wind are $\alpha$-particles, O$^{5+}$, and O$^{7+}$ ions \citep{Cranmer_2012}. The large relative masses of these heavy ions account for significant contributions to the overall ion mass and momentum densities despite their small relative abundance. Their signatures can be observed in measured ion velocity distribution functions \citep[VDFs;][]{Asbridge_1976,Marsch_1982a, DeMarco2023}. Key characteristics of these ion VDFs include field-aligned beams, temperature anisotropy with $T_{\perp}/T_{\parallel}>1$, where $T_{\perp}$ and $T_{\parallel}$ correspond to the temperatures of the distribution measured in the directions perpendicular or parallel to the local magnetic field, and skewness, which represents a non-zero field-aligned heat flux \citep{Verscharen_2019}. Heavy ions in the solar wind, such as $\alpha$-particles and oxygen ions, are also known to possess field-aligned drift velocities relative to proton populations. These drift velocities can be of the order of the local Alfv\'en speed in the weakly collisional fast solar wind \citep{Alterman_2018}, a smaller, but significant fraction of the Aflv\'en speed in the Alfv\'enic slow solar wind, or close to zero in the the non-Alfv\'enic slow solar \citep{Stansby_2020}.

The study of magnetized space plasmas has been intrinsically related to that of non-equilibrium thermodynamics and statistical physics \citep{Olbert_1968,Vasyliunas_1968,Cairns_1995, Livadiotis_2009, Ourabah_2020}. Many of these systems, including the solar wind and planetary magnetospheres, are weakly collisional due to the relative scarcity of Coulomb collisions between plasma particles along their respective trajectories; i.e.,  their mean free path is much greater than the system's characteristic spatial scales \citep{Marsch_1983,Marsch1991,Livadiotis_2019}. Therefore, the VDFs of the plasma particles are the result of both particle--particle and wave--particle interactions, leading to non-equilibrium structures in the distributions. Indeed, wave activity and its interactions with the plasma particles has been linked to the observed non-thermal features of proton and $\alpha$-particle VDFs since the early exploration of the solar wind  \citep{Feldman_1973, Marsch_1982b, Marsch_1982a}. 

Alfv\'enic fluctuations that propagate along the magnetic field increase the temperature anisotropy of protons and $\alpha$-particles through resonant interactions \citep{Gary2005, Bourouaine2011a}. The temperature anisotropy can drive wave instabilities with thresholds that depend on the kinetic-to-magnetic pressure ratio $\beta=8\pi nk_B T/B^2$, where $n$ is the particle number density, $k_B$ is the Boltzmann constant, $T$ is the plasma temperature, and $B$ is the strength of the local magnetic field. These instability thresholds depend on the plasma parameters for both protons \citep{Kasper2003, Hellinger2006} and $\alpha$-particles \citep{Maruca_2012,Li2023, McManus_2024}. 

Ion temperature ratios in the collisionally young solar wind exhibit an approximate mass proportionality with  $T^{(i)}/T^{(p)}\gtrsim m^{(i)}/m^{(p)}$, where $m^{(i)}\,\,(T^{(i)})$  and $m^{(p)}\,\,T^{(p)}$ denote the mass (temperature)of ions and protons quantities, respectively, and  \citep{Tracy2016}. Resonant interaction of $\alpha$-particles and minor ions with linear waves of the Alfv\'en/ion-cyclotron (A/IC) branch can effectively heat heavy ions preferentially due to their cyclotron frequencies being smaller than that of the protons \citep{Dusenbery1981,McKenzie1981,Marsch1982c,Isenberg1983, Isenberg1984a, Isenberg1984b,Gomberoff1991,Isenberg_2009,Navarro_2020}. 

The presence of A/IC waves can affect not only the thermodynamic properties of the ions but also the shape of their VDFs \citep{zhang2024}. Quasilinear theory of resonant wave--particle interactions predicts diffusion of resonant ions along constant energy shells in the wave's rest frame \citep{Cranmer2001, Marsch_Tu_2001, He_2015,Bowen_2022, Shuster2024}, such that the VDF  reaches a steady state when it is locally in velocity space a function of the wave-frame kinetic energy only \citep{Isenberg2001a, Isenberg2001b,Tu_2002}. This equilibrium state, which we define as a wave--particle equilibrium, is characterized by a minimization of the interaction between the wave and the particles such that the plasma becomes transparent to the wave. This transparency coincides with a suppression of wave damping such that $\gamma=0$; i.e., the wave reaches a marginally stable state. 

For all dispersive waves, the nonthermal features of the wave--particle equilibria are local in velocity space, depending on the resonance condition involving the particle velocity, the wave frequency, and the wave-vector. 

Heavy ions can greatly affect the dispersion relation and the properties of kinetic plasma waves \citep{Isenberg1984b, Gomberoff-1982,Moya_2022,Villarroel2023}. Even for ion-to-proton density ratios as small as $n_i/n^{(p)} \sim 10^{-4}$,  temperature anisotropies and velocity drifts with values observed in the solar wind provide an effective source of free energy for kinetic instabilities \citep{Gomberoff_Valdivia_2002, Gomberoff_Valdivia_2003, Moya_2014, McManus_2024,martinovic2026}. 

\section{Vlasov--Maxwell theory}
The plasma particles of species $s$  are statistically described by the VDF $f_s(t,\vec{r},\vec{v})$. These distribution functions satisfy the Vlasov equation in a weakly collisional plasma:
\begin{equation}
    \frac{\partial f_{s}}{\partial t}+\vec{v}\cdot\nabla_{\vec{r}}f_{s}+\frac{{q^{(s)}}}{m^{(s)}}\left[\vec{E}+\frac{\vec{v}}{c}\times\vec{B}\right]\cdot\nabla_{\vec{v}}f_s=0, \label{Eq: Vlasov}
\end{equation}
where $\nabla_{\vec{r}}$ is the spatial gradient, $\nabla_{\vec{v}}$ is the velocity-space gradient, $c$ is the speed of light in vacuum, and $q^{(s)}$ is the mass and electric charge of a particle of species $s$.
The system is coupled with Maxwell's equations through the charge density
\begin{equation}
\rho(t,\vec{r}) =\sum^{(s)}q^{(s)}\int f_s(t,\vec{r},\vec{v})d^3v, \label{eq:zero-charge}
\end{equation}
and the current density
\begin{equation}
    \vec{j}(t,\vec{r}) =\sum^{(s)}q^{(s)}\int \vec{v}f_{s}(t,\vec{r},\vec)d^3v.\label{eq:zero-current}
\end{equation}

We linearize this system of equations, such that the VDF is expressed as
\begin{equation}
    f_{s}(\vec{r},\vec{v},t)=f_{0}^{(s)}(\vec{v})+\delta f_{s}(\vec{r},\vec{v},t).
\end{equation}
Also, we consider a magnetized system with no background electric field, such that the magnetic field is linearized as
\begin{equation}
    \vec{B}(\vec{r},t)=\vec{B}_0 + \delta\vec{B}(\vec{r},t)
\end{equation}
and the electric field as
\begin{equation}
    \vec{E}(\vec{r},t)=\delta\vec{E}(\vec{r},t).
\end{equation}

\subsection{Dispersion relation of electromagnetic waves}
We define a cylindrical coordinate system in velocity space aligned with the background magnetic field $\vec B_0 =B_0\hat{\vec{z}}$. Velocity-space vectors are then expressed as $\vec{v}=v_{\perp}\hat{\vec{e}}_{\perp}+v_{\parallel}\hat{\vec{z}}$, and $\phi$  denotes the azimuthal angle of the velocity vector.
We assume that  ${\partial f_{0}^{(s)}}/{\partial {\phi}}=0$, which restricts the background distribution to be gyrotropic. 

The dispersion relation of electromagnetic waves is given by \citet{stix1992} as
\begin{equation}\vec{k}\times\left(\vec{k}\times\delta\vec{\hat{E}}({\vec{k}},\omega)\right)+\frac{\omega^2}{c^2}\left(\mathbf 1+\sum^{(s)} {\bm{\chi}}^{(s)}(\vec{k},\omega)\right)\cdot\delta\vec{\hat{E}}({\vec{k}}.\omega)=0, \label{disp}
\end{equation}
where  $\vec{k}$ is the wave-vector, $\omega=\omega_{\mathrm r}+i\gamma$ is the complex frequency with $\omega_{\mathrm{r}},\gamma\in\mathbb{R}$, and $\delta\hat{\vec{E}}(\vec{k},\omega)$ is the Fourier transform of $\delta\vec{E}(\vec{r},t)$. The imaginary part of the frequency, $\gamma$, is related to the temporal evolution of the wave's amplitude; the wave is damped when $\gamma<0$ and unstable when $\gamma>0$.  The electric susceptibility in the Fourier representation, ${\bm{\chi}}^{(s)}(\vec{k},\omega)$, is calculated from the linearized VDF and relates to the linearized electric field through Eq.~\eqref{eq:zero-current}.

When $|\gamma|\ll|\omega_{r}|$, the power transferred from the wave to the particles of species $s$ is given by \citep{stix1992}
\begin{align}
    P^{(s)}&=\frac{\omega_{\mathrm r}}{8\pi}\delta\mathbf{\hat{E}}^*({\vec{k},\omega)}\cdot \bm{\chi}^{(s)}_{(A)}(\vec{k},\omega)\cdot\delta\mathbf{E}({\vec{k}},\omega),
\end{align}
where $\bm{\chi}^{(s)}_{(A)}=(\bm{\chi}^{(s)}-\bm{\chi}^{(s)\dagger})/(2i)$ is the anti-Hermitian part of the electric susceptibility and $\bm{\chi}^{(s)\dagger}$ is the Hermitian transpose of $\bm{\chi}^{(s)}$.

\subsection{Quasilinear diffusion}
Considering the terms up to second order in the perturbation of the Vlasov--Maxwell system, we obtain the quasilinear evolution of the background distribution \citep{Shapiro_1963,Kennel_1966}. Then,
\begin{equation}
    \frac{\partial f_{0}^{(s)}}{\partial t}=i\left(\frac{q^{(s)}}{ m^{(s)}}\right)^2\sum_{\ell=-\infty}^{\infty}\int d\vec{k}\left[\frac{\partial}{\partial \alpha^*}\frac{\left|J_{\ell+1}(\lambda_s)\delta \hat{E}^{+}(\vec{k},\omega)+J_{\ell-1}(\lambda_s)\delta \hat{E}^{-}(\vec{k},\omega)\right|^2}{2(\omega-\ell\Omega^{(s)}-k_{\parallel}v_{\parallel})}\frac{\partial}{\partial \alpha}+\frac{\partial}{\partial \mu^*}\frac{J_\ell(\lambda_s)^2|\delta \hat{E}^z(\vec{k},\omega)|^2}{(\omega-\ell\Omega^{(s)}-k_{\parallel}v_{\parallel})}\frac{\partial}{\partial \mu}\right]f_{0}^{(s)},\label{eq:QL}
\end{equation}
where 
\begin{equation}
\frac{\partial}{\partial \alpha} = \left(\frac{k_{\parallel}v_{\parallel}}{\omega}-1\right)\frac{\partial}{\partial v_{\perp}}-\frac{k_{\parallel}v_{\perp}}{\omega}\frac{\partial}{\partial v_{\parallel}}\end{equation}
is the pitch angle gradient, 
\begin{equation}
\frac{\partial}{\partial \mu}=\left(1-\frac{\ell\Omega^{(s)}}{\omega}\right)\frac{\partial }{\partial v_{\parallel}}+\frac{\ell\Omega^{(s)}}{\omega}\frac{v_{\parallel}}{v_{\perp}}\frac{\partial }{\partial v_{\perp}},
\end{equation}
$\lambda_s={k_{\perp}v_{\perp}}/{\Omega^{(s)}}$, $\Omega^{(s)}=q^{(s)}B_0/m^{(s)} c$ is the species' gyrofrequency, and $\delta \hat{E}^{\pm}(\vec{k},\omega)=[\delta \hat{E}^{x}(\vec{k},\omega)\mp i\delta \hat{E}^{y}(\vec{k},\omega)]/\sqrt{2}$ \citep{Yoon2017}. 

Waves are in resonance with particles when the condition 
\begin{equation}
 k_{\parallel}v_{\parallel}=\omega_{\mathrm r}-\ell\Omega^{(s)} \label{eq:resonance}
\end{equation}
is fulfilled, where $\ell\in\mathbb{Z}$ and $\omega_{\mathrm{r}}=\text{Re}\{\omega\}$. Equation~\eqref{eq:QL} represents a diffusion equation for the background distribution function $f_{0}^{(s)}$ in the variables $\alpha$ and $\mu$. Resonant wave--particle interactions evolve the background VDF until the term in square brackets in Equation~\eqref{eq:QL} vanishes. For parallel propagation, this condition is fulfilled when $\partial f_{0}^{(s)}/\partial \alpha=0$ for purely electromagnetic waves, or when $\partial f_{s0}/\partial \mu=0$ for purely electrostatic waves. 

The dominant terms in the sum in Eq.~\eqref{eq:QL} are those involving $J_0(\lambda_s)$, especially when $\lambda_s\lesssim 1$. The first term on the right-hand side of Equation~\eqref{eq:QL} dominates for first-order cyclotron resonances $k_{\parallel}v_{\parallel}=\omega_{\mathrm r}\pm\Omega^{(s)}$. Particles that fulfill the resonance condition in Equation~\eqref{eq:resonance}  diffuse along shells of constant energy in the reference frame that is co-moving with the wave, i.e., locally tangent to curves defined by $v_{\perp}^2+(v_{\parallel}-\omega/k_{\parallel})^2=\text{constant}$ \citep{Besse_2011, Yoon2017}. The second term on the right-hand side of Equation~\eqref{eq:QL} dominates for the Landau resonance $k_{\parallel}v_{\parallel}=\omega_{\mathrm r}$, which evolves the background VDF towards one independent of $v_{\parallel}$ for $\ell=0$ in the vicinity of the resonant velocity.

\section{Test-particle simulations} \label{Simulation}
We simulate heavy ions as test-particles initially in thermal equilibrium to study the effects of resonant wave--particle interactions on heavy ions. The choice of considering the heavy ions as test-particles allows us to study the wave--particle interactions in a controlled fashion, as well as to quantify the effect of these species in the dispersion relation of the waves. We force the system with wave solutions based on the Vlasov--Maxwell dispersion relation. 

We study the evolution of the VDF of heavy ions towards their wave--particle equilibrium state by analyzing both the heavy ion VDF, its thermodynamic properties, and the dispersion relation of the involved waves using the Arbitrary Linear Plasma Solver (ALPS) \citep{Verscharen_2018,alps_2023_8075682}. The imprint of wave--particle interactions can drive plasma VDFs far from thermal equilibrium in ways that deviate strongly from idealized, bi-Maxwellian distributions. These nonthermal features, observed in kinetic simulations and \textit{in situ} observations, have important effects on the propagation of kinetic electromagnetic waves that cannot be captured by bi-Maxwellian solvers \citep{Walters_2023,McManus_2024,Klein_2025}.  The recently available ALPS code is able to calculate the dispersion relation of linear waves in plasmas with arbitrary VDFs, making it an ideal tool for studying wave propagation in the nonthermal wave--particle equilibrium states.

We consider a plasma consisting of electrons, protons, and a second species of ions. We assume that $n^{(p)}\gg n_{i}$ and $n^{(p)}/n_{e}\sim 1$. The quasi-neutrality condition demands $n_e=n^{(p)}+(q_i/e) n_i$, where $e$ is the elementary charge. 

We use the Boris algorithm to evolve the velocity and position of the test-particles \citep{Qin_2013,Tajima_2004}. 

Velocities are normalized to the proton Alfv\'en speed, $V_{A}^{(p)}=B_0/\sqrt{4\pi n^{(p)}m^{(p)}}$, lengths to the proton inertial length, $\ell^{(p)}=c/\omega_{p}^{(p)}$, where $\omega_{p}^{(p)}=\sqrt{4\pi n^{(p)} e^2/m^{(p)}}$,  time  to the proton gyroperiod, $\tau^{(p)}=1/\Omega^{(p)}$, and electromagnetic fields to the magnitude of the background magnetic field, $B_0$.

We initialize the system with a Maxwellian VDF for the test-particles using the Box-Mu  ller algorithm \citep{Box_1958}. The perturbed electromagnetic fields are fully kinetic solutions of the three-species dispersion relation with Maxwellian electrons, protons, and minor ions, obtained with ALPS, using the same initial parameters as the simulation. The system is spatially confined within a periodic box, centered at the origin, of size $2\pi \omega_{p}^{(p)}/ck_{\perp}$ along the $X$ and $Y$ directions, and $2\pi\omega_{p}^{(p)}/ck_{\parallel}$ along the $Z$ direction.  The initial spatial distribution of test-particles is uniform inside the box. Particles that leave the box on one side re-enter the box from the opposite side periodically and with the same velocity. 

We write the electric field as a monochromatic plane wave
\begin{equation}
    \delta \vec{E}(\vec{r},t)=\text{Re}\left\{e^{i({\mathbf{k}}\cdot\mathbf{{r}}-{\omega} {t})+\phi_0}\left(\delta \hat{E}_x(\vec{k},\omega)\hat{\vec{x}}+\delta \hat{E}_y(\vec{k},\omega)\hat{\vec{y}}+\delta \hat{E}_z(\vec{k},\omega)\hat{\vec{z}}\right)\right\},\label{eq:EF}
\end{equation}
where we obtain $\omega$ and $\delta \vec{\hat{E}}(\vec{k},\omega)\in\mathcal{C}$ from ALPS, and $\phi_0$ is a random phase.  
We calculate the magnetic field of the wave as
\begin{equation}
    \delta \vec{B}(\vec{r},t)=\text{Re}\left\{e^{i({\mathbf{k}}\cdot\mathbf{{r}}-{\omega} {t})+\phi_0}\left(\delta \hat{B}_x(\vec{k},\omega)\hat{\vec{x}}+\delta \hat{B}_y(\vec{k},\omega)\hat{\vec{y}}+\delta \hat{B}_z(\vec{k},\omega)\hat{\vec{z}}\right)\right\}.\label{eq:MF}
\end{equation}
Equations~\eqref{eq:EF} and \eqref{eq:MF} are related through Faraday's law $c\vec{k}/\omega \times \delta \hat{\vec{E}}=\delta\hat{\vec{B}}$, which connects the amplitudes $\delta \hat{\vec E}$ and $\delta \hat{\vec B}$. We set the magnitude of the forcing electromagnetic wave through a pre-defined choice of $|\delta B_y(t=0)|/B_0$. 

In each time step of our simulation, we derive the VDF by binning our test-particles in a two-dimensional grid in velocity space. 
We compute the bulk velocity as
\begin{align}
\vec{U}=\frac{1}{N}\sum_{n=1}^{N} \vec{v}_n,
\end{align}
where $\vec{v}_n$ is the velocity vector of the $n$-th test-particle in our sample, and $N$ is the total number of test-particles. We set $N=2\times10^{6}$ in all our simulations.

We derive the parallel temperature as the kinetic energy of the particles in the Lagrangian reference frame, 
\begin{equation}
    T_{\parallel}^{(s)}=\frac{m^{(s)}}{k_{B}N}\sum_{n=1}^{N}({{v}_{n,z}}-{{U}_z})^2, \label{eq: tpar}
\end{equation}
Similarly, we compute the perpendicular temperature as
\begin{equation}
    T_{\perp}^{(s)}=\frac{m^{(s)}}{k_BN}\sum_{n=1}^{N}\frac{1}{2}\left[({{v}_{n,x}}-{{U}_x})^2+({{v}_{n,y}}-{{U}_y})^2\right]. \label{eq: tperp}
\end{equation}
We derive the heat flux as the third velocity-moment of the distribution:
\begin{equation}
    \vec{q}^{(s)} =\frac{1}{2}\frac{m^{(s)}}{N}\sum_{n=1}^{N}|{\vec{v}}_{n}-{\vec{U}}|^2({\vec{v}_n}-{\vec{U}}).\label{eq:heat}
\end{equation}

\section{Results}
We present the results of the simulation and solutions to the dispersion relation in the following subsections.

\subsection{Linear electromagnetic waves in a warm Maxwellian multi-ion plasma}
The system is initially described by Maxwellian VDFs for all plasma species. We solve Eq.~\eqref{disp} numerically with ALPS to find the electromagnetic fields of a given kinetic plasma wave, which we then use to perturb the system through the mechanism described in Section~\ref{Simulation}. For this, we set $V_{A}^{(p)}/c=10^{-4}$, $T^{(e)}=T^{(p)}$, and $T^{(i)}=(m^{(i)}/m^{(p)})T^{(p)}$. Incorporating the quasineutrality condition, and normalizing to proton quantities, this leads to the input parameters being related by $n^{(e)}/n^{(p)}=1+q^{(i)}n^{(i)}/en^{(p)}$, $\beta^{(e)}=n^{(e)}\beta^{(p)}/n^{(p)}$, and $\beta^{(i)}=n^{(i)}m^{(i)}\beta^{(p)}/n^{(p)}m^{(p)}$. This is equivalent to setting $v_{\text{th}}^{(i)}=v_{\text{th}}^{(p)}$, where $v_{\text{th}}=\left(2k_BT^{(s)}/m^{(s)}\right)^{1/2}$ is the thermal speed of species $s$, with $\beta^{(i,p)}=\left(v_{\text{th}}^{(i,p)}/V_{A}^{(p)}\right)^2$. In the following text, the number density and thermal speed for ions corresponding to the input parameters used in ALPS will be given. 
\begin{figure*}[h!]
        \centering
    \includegraphics[width=\textwidth]{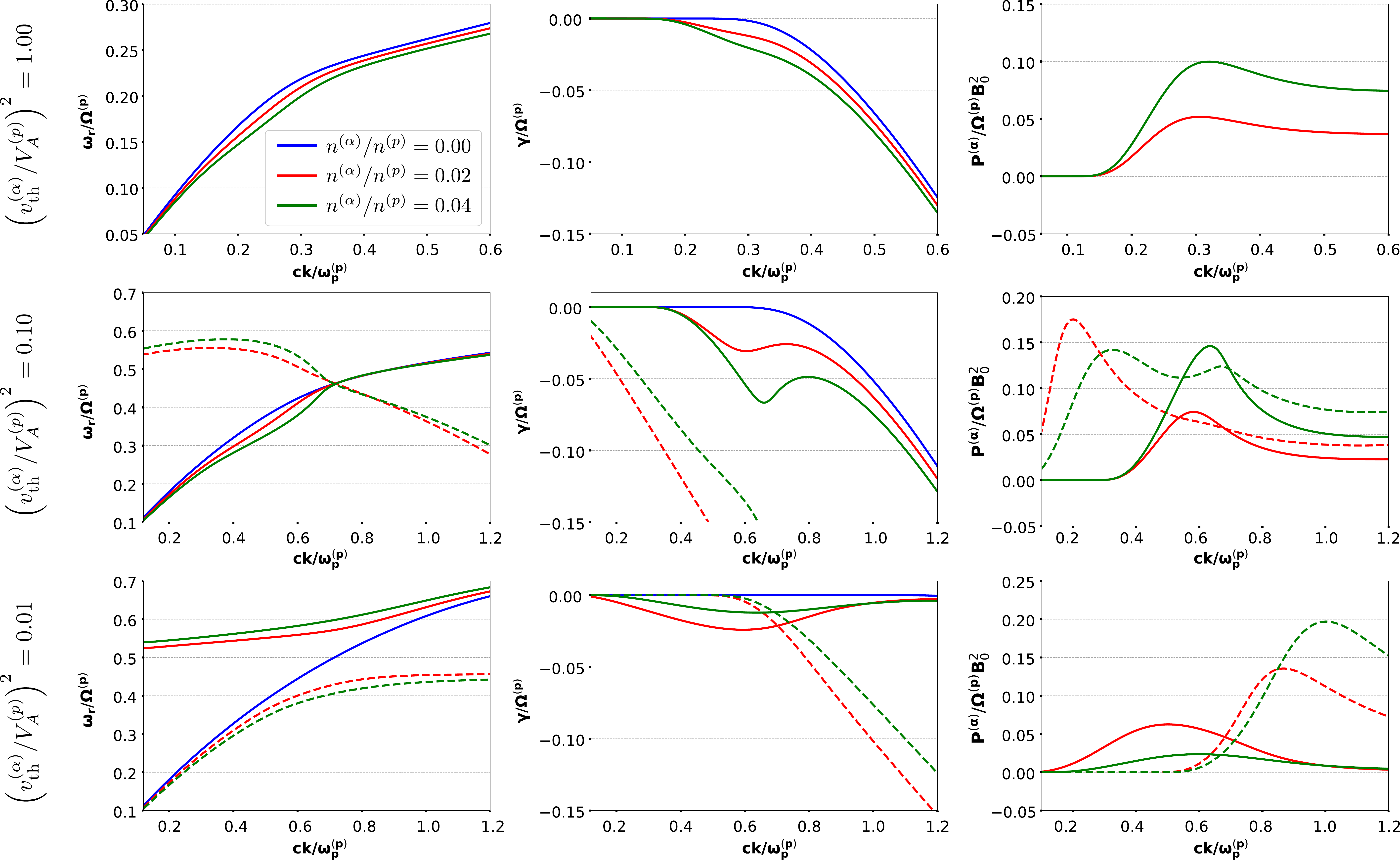}
    \caption{Real (left) and imaginary (center) parts of the frequency and collisionless heating rate (right) of A/IC waves with varying thermal speeds, given $T^{(\alpha)}/T^{(p)}=m^{(\alpha)}/m^{(p)}$ for different values of $v_{\text{th}}^{(\alpha)}$, and different number densities. When the solution splits into branches, we plot the one that is asymptotic to the A/IC solution at large wavenumbers with continuous lines, while we plot the remaining branch with dashed lines.  \label{fig:disp_alpha}}
\end{figure*}
Figure~\ref{fig:disp_alpha} displays the complex frequency and collisionless heating rate of parallel propagating A/IC waves in electron-proton and electron-proton-$\alpha$-particle plasmas with different thermal velocities. The inclusion of the $\alpha$-particles reduces the phase velocity of the wave and increases the damping rate of the wave. This relation between damping and resonant heating is made explicit by comparing the collisionless heating rate of $\alpha$-particles, $\text{P}_{\alpha}$, (right column) with the wave damping, $\gamma/\Omega^{(p)}$, (middle column), as these quantities are inversely correlated.

When the plasma is cold, i.e. $v_{\text{th}}^{(s)}/ V_{A}^{(p)}\ll1$, as in the bottom row of Figure~\ref{fig:disp_alpha}, which considers $\beta^{(p)}=0.01$, two distinct dispersion relations for A/IC waves appear in the presence of $\alpha$-particles. In general, in a cold Maxwellian plasma, A/IC waves exhibit asymptotes at the gyrofrequencies of the different ion species that comprise the plasma, thus splitting the dispersion relation into disjointed frequency bands. Here, the dashed solutions asymptote towards $\Omega^{(\alpha)}=0.5\Omega^{(p)}$, while the continuous solutions tend to $\omega_{\mathrm r}\to\Omega^{(p)}$ at large wavenumbers, which is consistent with the behavior of the electron-proton A/IC wave in the (semi-)cold approximation. With increasing temperature, the same reasoning applies to the protons, and therefore, the real part of the frequency must approach $\Omega^{(p)}$ as $k_{\parallel}\gg1$ in a hot plasma. 

Because we consider $n^{(s)}\ll n^{(p)}$, the splitting of the solutions only appears when the plasma is cold. When the plasma is warm, as in the middle row of Figure~\ref{fig:disp_alpha}, the two solutions intersect. Here, the branch that results from the continuous extension of the Alfv\'en wave into the large-wavenumber domain resembles the electron-proton A/IC wave, while the second branch is strongly damped and acquires negative phase velocity at $kc/\omega_{p}^{(p)}\sim0.5$. \cite{Isenberg1984a} provides a detailed discussion on the nature of the two branches of the dispersion relation in an electron-proton-$\alpha$-particle plasma and on wave propagation for $\omega_{\mathrm r}\to\Omega^{(\alpha)}$.
\begin{figure*}[h!]
        \centering
    \includegraphics[width=\textwidth]{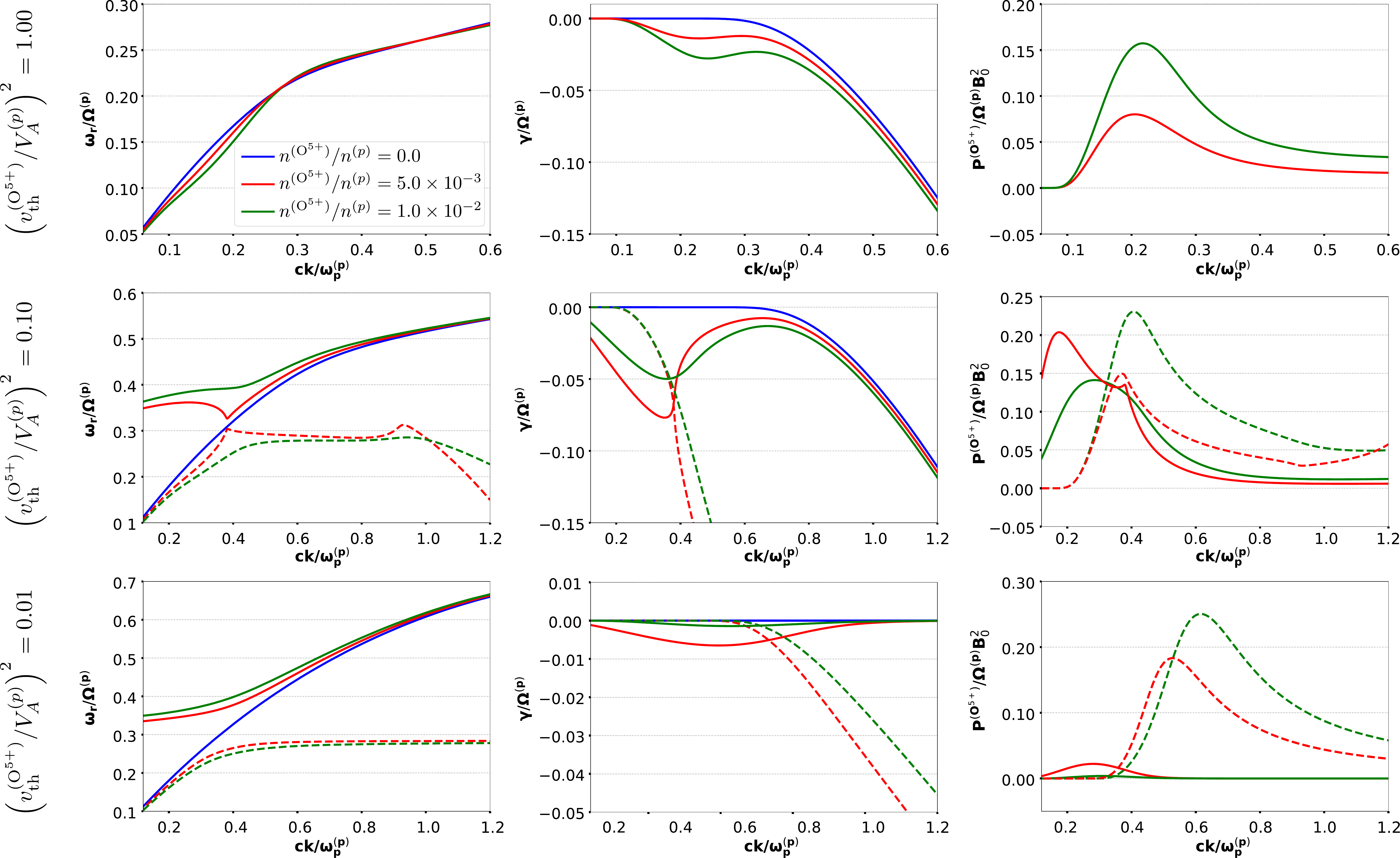}
    \caption{Real (left) and imaginary (center) parts of the frequency and collisionless heating rate (right) of A/IC waves in an electron-proton-O$^{5+}$ with varying thermal speeds, given $T^{(\text{O}^{5+})}/T^{(p)}=m^{(\text{O}^{5+})}/m^{(p)}$ for different values of $\beta^{(p)}$, and different number densities.  \label{fig:disp_O5} }
 \end{figure*}
Figure~\ref{fig:disp_O5} is analogous to Figure~\ref{fig:disp_alpha}, but for the case of an electron-proton-O${}^{5+}$ plasma. The considered heavy ion densities are chosen such that $\beta^{(\text{O}^{5+})}=\beta^{(\alpha)}$ for all equivalent curves in Figures~\ref{fig:disp_alpha} and \ref{fig:disp_O5}, given our mass-proportional temperature assumption. 
 
\begin{table}[h!]
  \centering
\setlength{\arrayrulewidth}{0.5mm}
\begin{tabular}{|c|c|c|c|c|c|c|}
\hline
\multicolumn{7}{|c|}{\centering Initial normalised plasma and wave parameters}\\
\hline
\hline
    Thermal speed&Ion species & Drift velocity ($V_D$) & $n^{(s)}/n^{(p)}$& $ck_0/\omega_{p}^{(p)}$ & $\text{Re}\{\omega_0\}/\Omega^{(p)}$ & $\text{Im}\{\omega_0\}/\Omega^{(p)}$ \\
    \hline
  \multirow{5}{3cm}{$\left(v_{\text{th}}^{(s)}/V_{A}^{(p)}\right)^2=1.0$} & $\alpha$ & $V_D=0.0$ &$0.040$ & $0.300$ & $0.200$ & $-0.021$   \\
  & $\alpha$ & $V_D=V_{A}^{(p)}$& $0.040$ & $0.300$ & $0.219$ & $-0.001$   \\
  & $\alpha$ & $V_D=-V_{A}^{(p)}$ &$0.040$ & $0.300$ & $0.347$ & $-0.091$   \\
   & O$^{7+}$ & $V_D=0.0
   $ &$0.010$ & $0.300$ & $0.205$ & $-0.025$   \\
   & O$^{5+}$ & $V_D=0.0$ &$0.010$& $0.300$ & $0.221$ & $-0.024$  \\
   \hline 
   \multirow{6}{3cm}{$\left(v_{\text{th}}^{(s)}/V_{A}^{(p)}\right)^2=0.1$} &  $\alpha$ &$V_D=0.0$ & $0.040$ & $0.660$ & $0.425$ & $-0.067$  \\
   & O$^{7+}$ & $V_D=0.0$ &$0.010$ & $0.580$ & $0.377$ & $-0.09$  \\
   & O$^{5+}$ & $V_D=0.0$ &$0.010$ & $0.367$ & $0.238$ & $-0.050$  \\
    &  O$^{5+}$ &$V_D=0.0$ & $0.010$ & $0.367$ & $0.391$ & $-0.050$   \\
    &  O$^{5+}$ & $V_D=0.0$ &$0.005$ & $0.380$ & $0.302$ & $-0.071$  \\
    &  O$^{5+}$ & $V_D=0.0$ &$0.005$ & $0.380$ & $0.327$ & $-0.065$   \\
    \hline
\end{tabular}
    \caption{Plasma parameters and selected dispersion relations used in the test-particle simulations.\label{tab:Table1}}
    
\end{table}
 The values of the relevant parameters employed for the simulations are collected in Table~\ref{tab:Table1}. Here, the drift velocity corresponds to the field aligned relative drifting speed of the ion population relative to the protons. For the initial parameters of the simulation, we chose wavenumbers close to the local minimum in the damping rate, which are those for which energy transfer from the electromagnetic waves to the initially Maxwellian particles is most efficient. The fraction of resonant particles is 
\begin{equation}
\Delta N/N \propto \exp\left(-\frac{\left(\omega_{\mathrm r}-\Omega^{(s)}\right)^2}{\left(k_{\parallel}v_{th \parallel}^{(s)}\right)^2}\right).\label{eq:deltaN}
\end{equation}
Replacement of the wavenumber values in Table~\ref{tab:Table1} and their corresponding frequencies into Equation~\eqref{eq:deltaN} yields comparable numbers for all driftless cases. 

Because the ions are simulated as test-particles, they possess no influence on the evolution of the electromagnetic fields, whose squared amplitude evolves according to the linear theory time dependence $|B(t)|^2=|B(t=0)|^2e^{2\gamma t}$. We perform the simulations for time periods given by $T=2\pi/|\gamma|$, which ensures that the wave amplitudes are damped to $\sim0.2\%$ of their initial value. 

\subsection{Particle kinetics of cyclotron resonance}
We simulate the interaction between A/IC waves and heavy ions using different concentrations of $\alpha$-particles, O${}^{5+}$, and O${}^{7+}$ ions. We perform the test-particle simulation until the wave amplitude has approached zero due to damping and the system has become stationary.  
\begin{figure*}[h!]
        \centering
    \includegraphics[width=\textwidth]{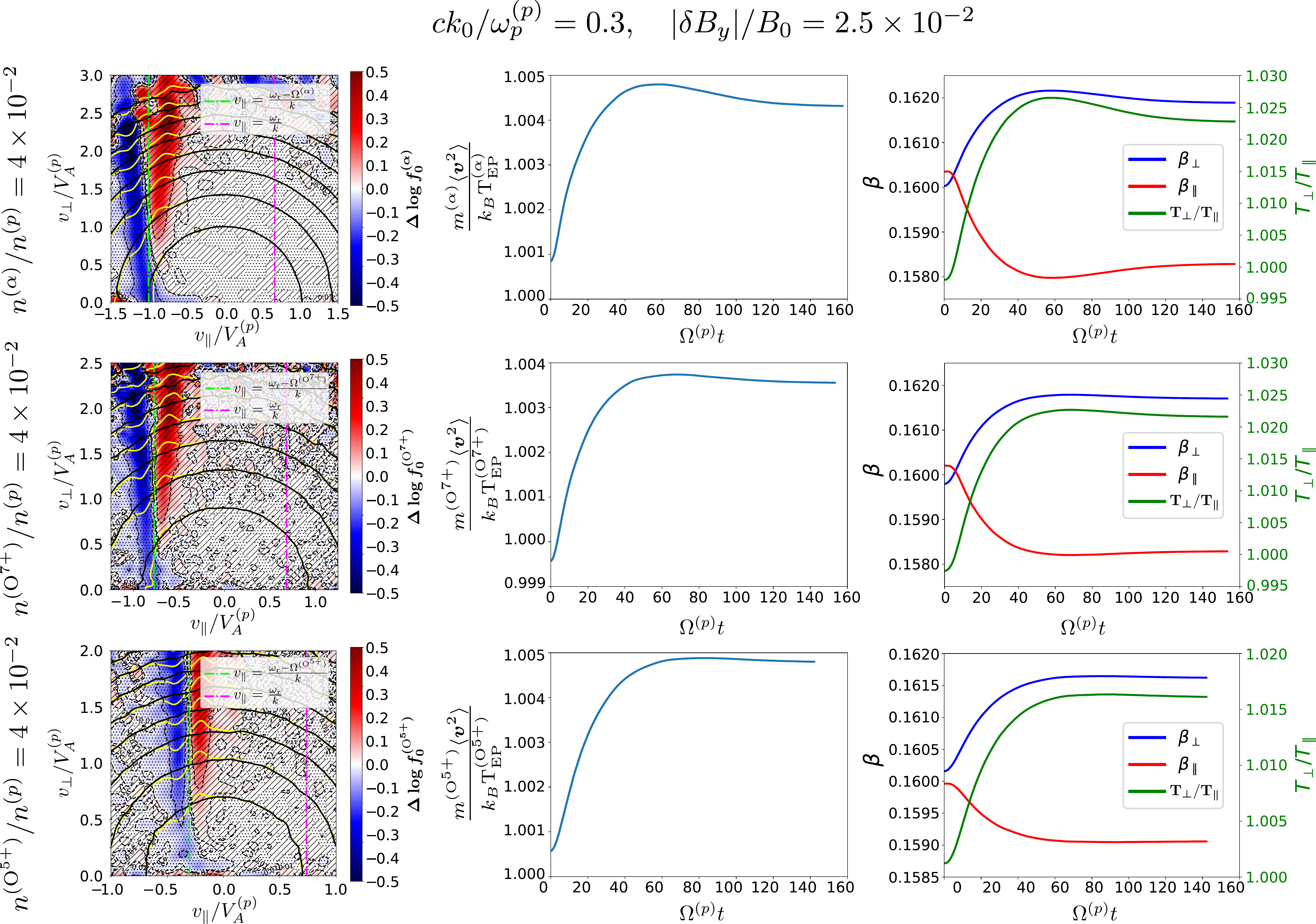}
    \caption{Results of the simulation for $\alpha$-particles (top panel), O$^{7+}$ ions (middle panel), and  O$^{5+}$ ions (bottom panel), with initial thermal speeds of $\left(v_{\text{th}}^{(s)}/V_{A}^{(p)}\right)^2 =1.0$. We choose the number densities such that $\beta^{(s)}(t=0)=0.16$. The forcing wave has a normalized wavenumber of $ck/\omega_{p}^{(p)}=0.3$ and an amplitude of $|\delta B_y|(t=0)/B_0=2.5\times10^{-2}$. The left column displays the logarithmic difference $\Delta \log(f_0^{(s)})=\log(f_{\text{final}})-\log(f_{\text{initial}})$ of the test-particles' final and initial VDFs for heavy ions; continuous contours are isocontours of the initial (black) and final (yellow) VDFs, dashed contours indicate curves of constant $\Delta \log(f)=\pm0.01,\pm0.10$, and vertical dash-dotted lines indicate the $\ell=0$ and $\ell=1$ resonances in Equation~\eqref{eq:resonance}.  The center column shows the mean kinetic-to-thermal energy ratio of the test-particles as a function of time. The right column shows $\beta_{\perp}$ (blue) and $\beta_{\parallel}$ (red), as well as the temperature anisotropy $T_{\perp}/T_{\parallel}$ (green) as functions of time.\label{fig:deltaf}}
    
\end{figure*}
Figure~\ref{fig:deltaf} displays the evolution of the VDFs, mean kinetic energy, $\beta_{\perp, \parallel}^{(s)}$, and $T_{\perp }^{(s)}/T_{\parallel}^{(s)} $, for $\alpha$-particles (top), O${}^{7+}$ (middle), and O${}^{5+}$ ions (bottom) with $\beta^{(s)}(t=0)=0.16$,  $v_{\text{th}}^{(s)}(t=0)/V_{A}^{(p)}=1.0$, and $T_{\perp s}(t=0)=T_{\parallel s}(t=0)$. The forcing waves possess wavenumbers of $ck/\omega_{p}^{(p)}=0.3$ and an initial amplitude of $|\delta B_y|/B_0=2.5\times10^{-2}$. 

The left column shows $\Delta \log f = \log f_{\text{final}}-\log f_{\text{initial}}$ for the three species of ions, where $f_{\text{initial}}$ is the initial, Maxwellian VDF and $f_{\text{final}}$ is the evolved VDF at the end of the simulation. We plot isocontours of the initial (final) VDFs in black (yellow), and use dashed lines to plot the resonant velocities which are solutions to Equation~\eqref{eq:resonance} with $\ell=0$ (magenta) and  $\ell=1$ (green). 

The VDF is modified near the resonant velocity in a way such that $f_{\text{final}}>f_{\text{initial}}$ for $v_{\parallel}>(\omega_{\mathrm r}-\Omega^{(s)})/k$ and $f_{\text{final}}<f_{\text{initial}}$ for $v_{\parallel}<(\omega_{\mathrm r}-\Omega^{(s)})/k$. 

The kinetic energy is normalized to that predicted by the equipartion theorem in ideal gases, which in normalized quantities is given by $k_BT_{\mathrm{EP}}^{(s)}/m^{(p)}V_{A}^{(p)2}=(3m^{(s)}/4m^{(p)})\left(v_{\text{th}}^{(s)}/V_{A}^{(p)}\right)^2$. Here, the thermal speed $v_{\mathrm{th}}^{(s)}$ corresponds to the one used for the initialization of the Maxwellian VDF used in the simulation and is not a dynamical quantity. In this particular case,  $\left(v_{\text{th}}^{(s)}/V_{A}^{(p)}\right)^2=1.0$ in this particular case. The test-particles gain kinetic energy overall, as can be appreciated in the middle column. For $\alpha$-particles, the kinetic-to-thermal energy ratio increases from $m^{(\alpha)}\langle\vec{v}^2\rangle/k_BT_{\mathrm{EP}}^{(\alpha)}\approx1.001$ toward a local maximum of approximately $m^{(\alpha)}\langle\vec{v}^2\rangle/k_BT_{\mathrm{EP}}^{(\alpha)}\approx1.005$ at $\Omega^{(p)}t\approx 50$, and then slightly decreases until $m^{(\alpha)}\langle\vec{v}^2\rangle/k_BT_{\mathrm{EP}}^{(\alpha)}\approx 1.004$. A similar trend is observed for oxygen ions. For $\mathrm{O}^{7+}$, the normalized kinetic energy begins at $\sim 0.9995$ and reaches a final value of $\sim 1.0035$, while for $\mathrm{O}^{5+}$, the energy increases from $\sim 1.0005$ to $\sim 1.005$. In both cases, a local maximum is also attained around $\Omega^{(p)}t \approx 50$.

Initially, all species fulfill $T_{\perp s}/T_{\parallel s}\sim 1$. This quantity behaves similar to the kinetic energy and acquires final values of $T_{\perp}^{(\alpha)}/T_{\parallel}^{(\alpha)}\sim1.023$, $T_{\perp}^{(\text{O}^{7+})}/T_{\parallel}^{(\text{O}^{7+})}\sim1.022$, and $T_{\perp}^{(\text{O}^{5+})}/T_{\parallel}^{(\text{O}^{5+})}\sim1.016$ for the respective species.  

For all species, $\beta_{\perp  s}(t=0)=\beta_{\parallel s}(t=0)\sim 0.16$. The plot shows that $\beta_{\perp}^{(s)}$ increases and $\beta_{\parallel}^{(s)}$ decreases with time. At the final time, $\beta_{\perp}^{(\alpha)}\sim0.1619$, $\beta_{\perp}^{(\alpha)}\sim0.1583$, $\beta_{\perp}^{(\text{O}^{7+})}\sim0.1615$, $\beta_{\parallel}^{(\text{O}^{7+})}\sim0.1585$, $\beta_{\perp}^{(\text{O}^{5+})}\sim0.1615$, and $\beta_{\perp}^{(\text{O}^{5+})}\sim0.1590$. 
\begin{figure*}[h!]
    \centering
    \includegraphics[width=0.88\textwidth]{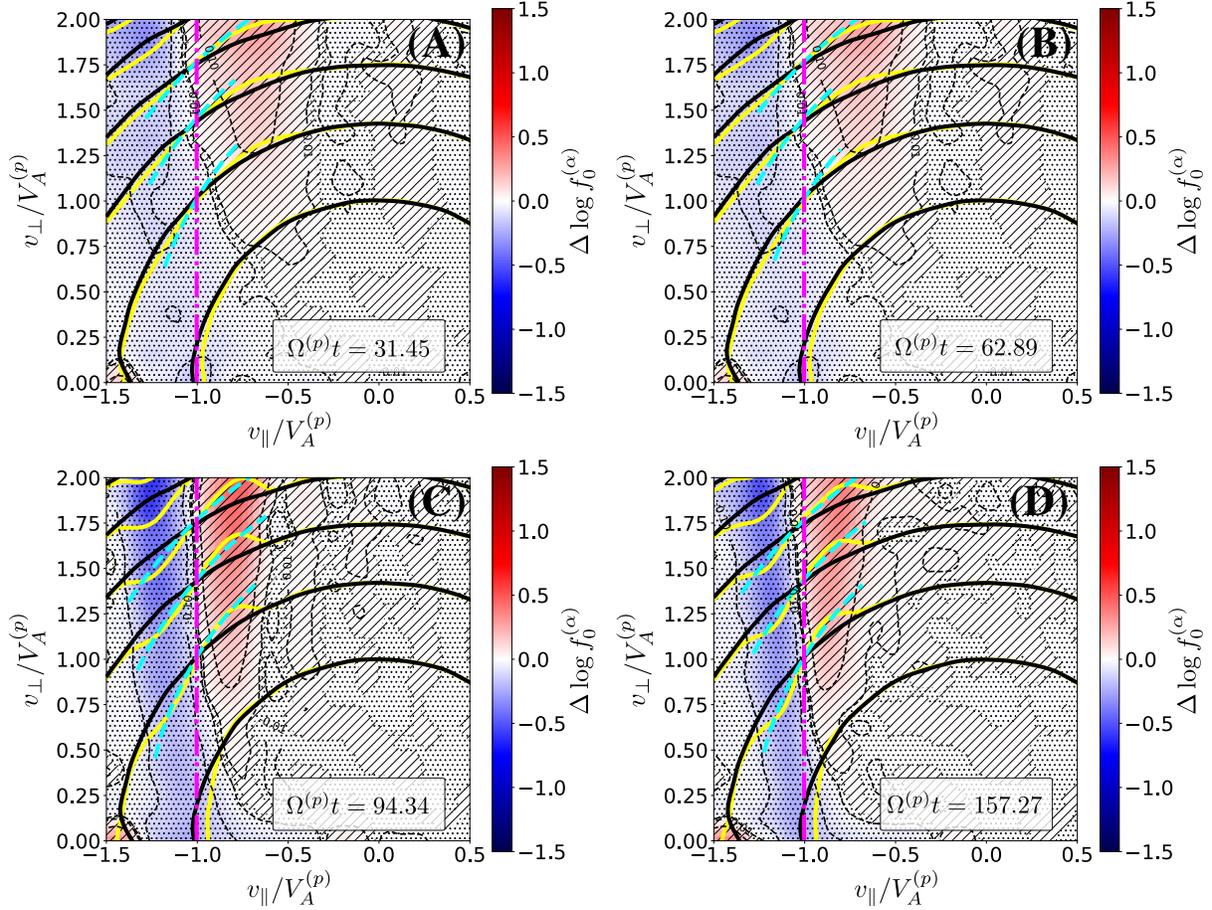}
    \caption{Logarithmic difference of the final and initial $\alpha$-particle VDFs near the cyclotron-resonant velocity for $ck/\omega_{p}^{(p)}=0.3$ and $|\delta B_y|/B_0 = 2.5\times 10^{-2}$ at different stages of the simulation. Contours of the initial and evolved VDFs are plotted in black and yellow lines, respectively. Dashed cyan lines represent shells of constant energy in the wave-frame, given by $(v_{\parallel}-\omega_{\mathrm r}/k_{\parallel})^2+v_{\perp}^2=\text{constant.}$ \label{fig: kin_shell}}
\end{figure*}
Figure~\ref{fig: kin_shell} shows a close-up of the logarithmic difference between the evolved and initial VDFs at different moments of the simulation in the vicinity of the cyclotron-resonant velocity for $\alpha-$particles with the same parameters as in the top panel of Figure~\ref{fig:deltaf}. The isocontours of the final distribution resemble constant energy contours in the reference frame that is co-moving with the wave, i.e., $(v_{\parallel}-\omega_{\mathrm r}/k)^2+v_{\perp}^2$, near the resonant velocity, $v_{\parallel}=(\omega_{\mathrm r}-\Omega^{(\alpha)})/k$, which are overplotted in cyan dashed lines. There are no significant differences between panel (C), that represents the VDF at $\Omega^{(p)}t=94.34$, and panel (D), that represents the final VDF at $\Omega^{(p)}t=157.27$, which is consistent with the evolution of the thermodynamic variables shown in Figure~\eqref{fig:deltaf}.

\subsection{Evolution of macroscopic thermodynamic variables}
As seen in Figures~\ref{fig:deltaf} and \ref{fig: kin_shell}, the resonant interaction of heavy ions with A/IC waves modifies the distributions from the initial Maxwellian. The resulting distributions are asymmetrical with respect to $v_{\parallel}/V_{A}^{(p)}=0$, as can be observed in the yellow isocontours in these Figures. This asymmetry is not present in the original Maxwellian VDFs and is developed over time until the system reaches a stationary state. This statistical skewness results in nonzero odd-order velocity moments of the VDF, which are zero in a  Maxwellian. 
\begin{figure*}
    \centering
    \includegraphics[width=\linewidth]{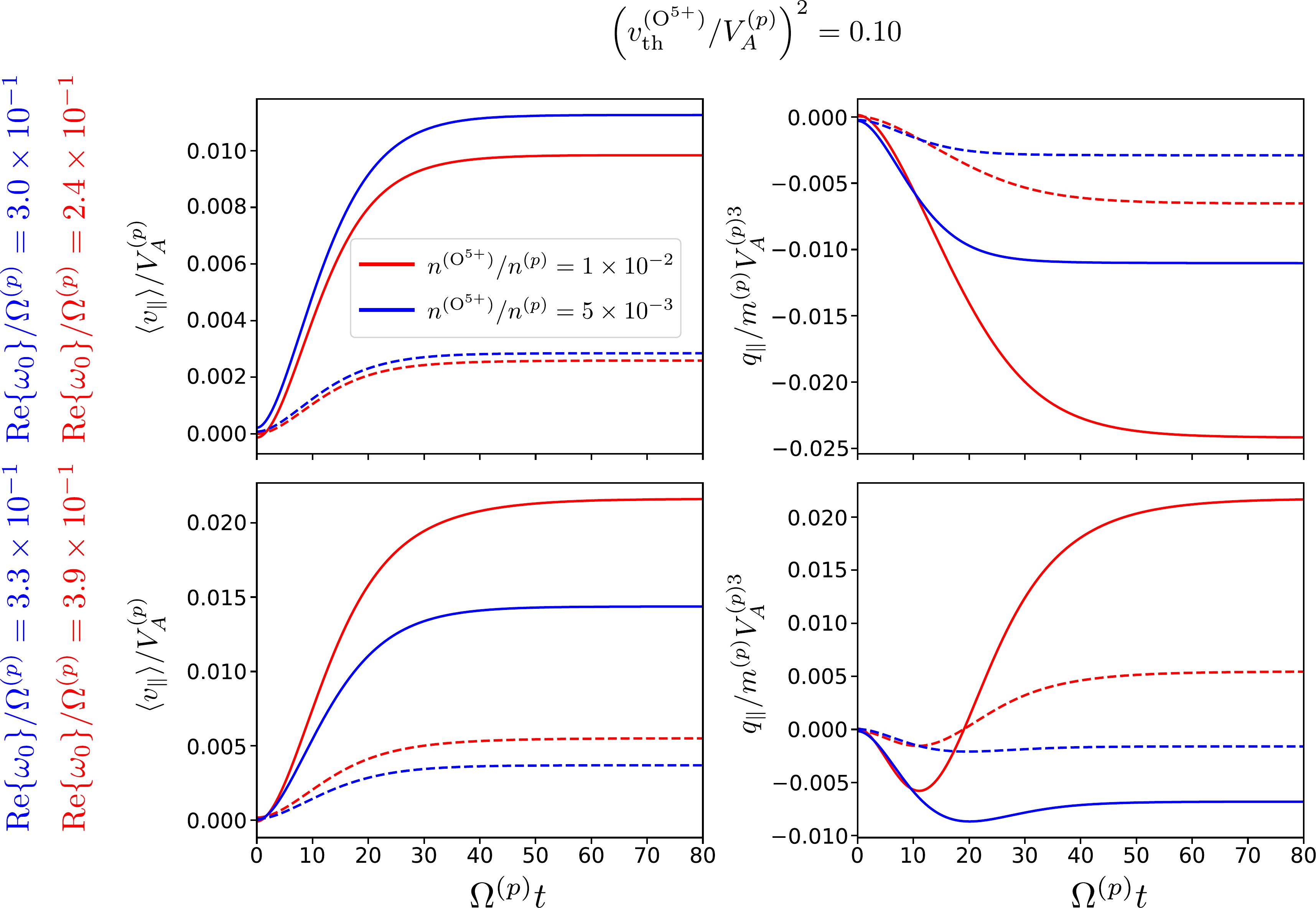}
    \caption{Mean field-aligned velocity and heat flux for $\left(v_{\text{th}}^{(\text{O}^{5+})}/V_{A}^{(p)}\right)^2=0.1$ for O${}^{5+}$ ions with different concentrations, forced by A/IC waves of the upper and lower branch. Continuous (dashed) curves indicate forcing by a wave with of amplitude $|\delta B_y|/B_0=5.0\times 10^{-2}$ ( $|\delta B_y|/B_0=2.5\times 10^{-2}$).\label{fig:heat}}
    
\end{figure*}
Figure~\ref{fig:heat} shows the time evolution of the odd-order moments of the VDF:  the field-parallel bulk velocity $\langle v_{\parallel}\rangle/V_{A}^{(p)}$ and the field-parallel heat flux $q_{\parallel}/m^{(p)} V_{A}^{(p)3}$ in an electron-proton-O${}^{5+}$ plasma with $\left(v_{\text{th}}^{(p)}/V_{A}^{(p)}\right)^2=\left(v_{\text{th}}^{(\text{O}^{5+})}/V_{A}^{(p)}\right)^2=0.1$, subject to the interaction with an A/IC wave. The wavenumber of the forcing wave is $ck/\omega_{p}^{(p)}=0.37$ for $n^{(\text{O}^{5+})}/n^{(p)}=1\times10^{-2}$, and $ck/\omega_{p}^{(p)}=0.38$ for $n^{(\text{O}^{5+})}/n^{(p)}=5\times10^{-3}$. The mean velocity aligned with the field and the heat flux are both initially zero in all cases. 

When the particles interact with the lower branch of the A/IC dispersion relation (top panels), for $n^{(\text{O}^{5+})}/n^{(p)}=1\times10^{-2}$ the bulk velocity evolves toward $\langle v_{\parallel}\rangle/V_{A}^{(p)}\approx2\times10^{-3}$ and the heat flux toward $q_{\parallel}/m^{(p)}V_{A}^{(p)3}\approx-6\times10^{-3}$ for $|\delta B_y(t=0)|/B_0=2\times10^{-2}$. When $|\delta B_y(t=0)|/B_0=5.0\times10^{-2}$, they evolve toward $\langle v_{\parallel}\rangle/V_{A}^{(p)}\approx9\times10^{-2}$ and $q_{\parallel}/m^{(p)}V_{A}^{(p)3}\approx-2.4\times10^{-2}$. For $n^{(\text{O}^{5+})}/n^{(p)}=5\times10^{-3}$, the final bulk speed is only slightly smaller than in the previous cases, while the heat flux is approximately half that of the equivalent cases with the highest O$^{5+}$ concentration, maintaining the negative sign.
When the forcing wave belongs to the upper branch of the A/IC dispersion relation (bottom panels), the time evolution and final values of the bulk speed and heat flux are similar to those discussed in the previous case when $n^{(\text{O}^{5+})}/n^{(p)}=5\times10^{-3}$. For $n^{(\text{O}^{5+})}/n^{(p)}=1\times10^{-2}$, the bulk speed values are approximately twice those obtained for the smaller ion concentration for the respective values of $|\delta B_y(t=0)|/B_0$. The heat flux behaves differently as it is initially negative, evolves towards a global minimum near $t\Omega^{(p)}\approx12$, and reverses its sign to become positive at $\Omega^{(p)}t\approx 20$, acquiring final values of approximately equal magnitudes to those obtained in the top right panel, but of opposite sign. 
 
\subsection{Dispersion relation and wave--particle equilibria}
We now use ALPS to solve the dispersion relation of parallel-propagating A/IC. For this, we used Maxwellian distributions for protons and electrons. For the heavy ion species, we used the actual VDF obtained from the stationary states of the test-particle simulations, and we set the electron drift speed as that required to satisfy the zero-current condition, given in the proton frame by $-n^{(e)}\boldsymbol{V}_D^{(e)}/n^{(p)}V_{A}^{(p)}+n^{(i)}q^{(i)}\boldsymbol{V}_D^{(i)}/n^{(p)}eV_{A}^{(p)}=0$, where $\boldsymbol{V}_{D}^{(s)}$ denotes the drift velocity of species $s$ relative to the protons. We employed the generalized linear least squares method to represent the analytical continuation of the VDF as a fit of Chebyshev polynomials of order 50, which ensures convergence of the results. We identify the solutions of the A/IC dispersion relation by checking that the electric fields of the waves satisfy $\mathrm{sign}(\omega_{\mathrm{r}})\left(\delta\hat{E}_x/i\delta\hat{E}_y\right)=-1$, and $\delta\hat{E}_z=0$, and qualitatively comparing with the dispersion relations obtained in Maxwellian plasmas.
\begin{figure*}[h!]
        \centering
    \includegraphics[width=\textwidth]{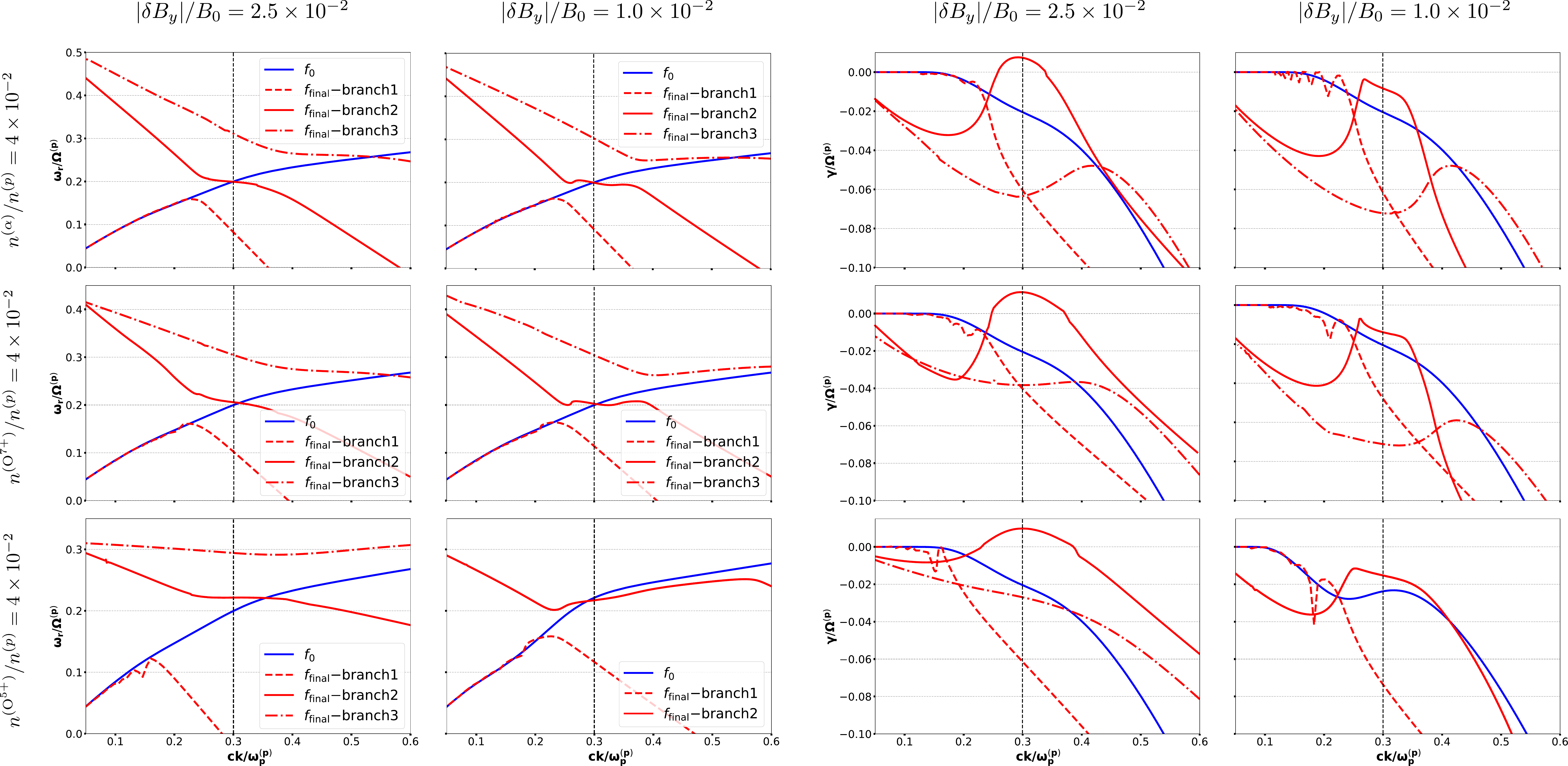}
    \caption{Real (left) and imaginary (right) parts of the dispersion relation of A/IC waves obtained with $\beta^{(s)}=1.6\times 10^{-1}$ for $\alpha$-particles, O${}^{5+}$, and O${}^{7+}$ ions. The results are obtained with the initial Maxwell distribution (blue) and final wave--particle quilibrium distribution (red) with different initial wave amplitudes.  \label{fig:real_w1}}
\end{figure*}
Figure~\ref{fig:real_w1} displays the dispersion relation of A/IC waves with the same parameters as in Figure~\ref{fig:deltaf}, considering waves with initial amplitudes of $|\delta B_y|/B_0=1.0\times10^{-2}$ and $|\delta B_y|/B_0=2.5\times10^{-2}$. The dispersion relations of the original Maxwellian plasma are plotted in blue, while those obtained with the new (final) distribution are plotted in red. Black, vertical dashed lines indicate the value of the wavenumber of the forcing wave, $k_0$. For this case, $ck_0/\omega_p^{(p)}=0.3$.

The results show that the $\omega_{\mathrm r}(k)$ splits into two or three frequency bands. We plot in continuous red lines the solutions for which $\omega_{\mathrm r}(k_0)$ is closest to that of the original forcing wave, $\omega_0(k_0)$. For the other frequency bands, we use dashed or dash-dotted red lines. At $k=k_0$, the real frequency varies only slightly with respect to $\omega_0(k_0)$. However, wave damping is reduced in all the cases considered. In particular, for $\alpha$-particles and O${}^{7+}$ ions, and a forcing wave with an amplitude of $|\delta B_y|/B_0=2.5\times 10^{-2}$, the waves become unstable at $k_0$, with values reaching $\gamma/\Omega^{(p)}\approx1.0\times10^{-2}$ for $\alpha$-particles and $\gamma/\Omega^{(p)}\approx1.5\times10^{-2}$ for Oxygen ions. 
\begin{figure*}[h!]
        \centering
    \includegraphics[width=\textwidth]{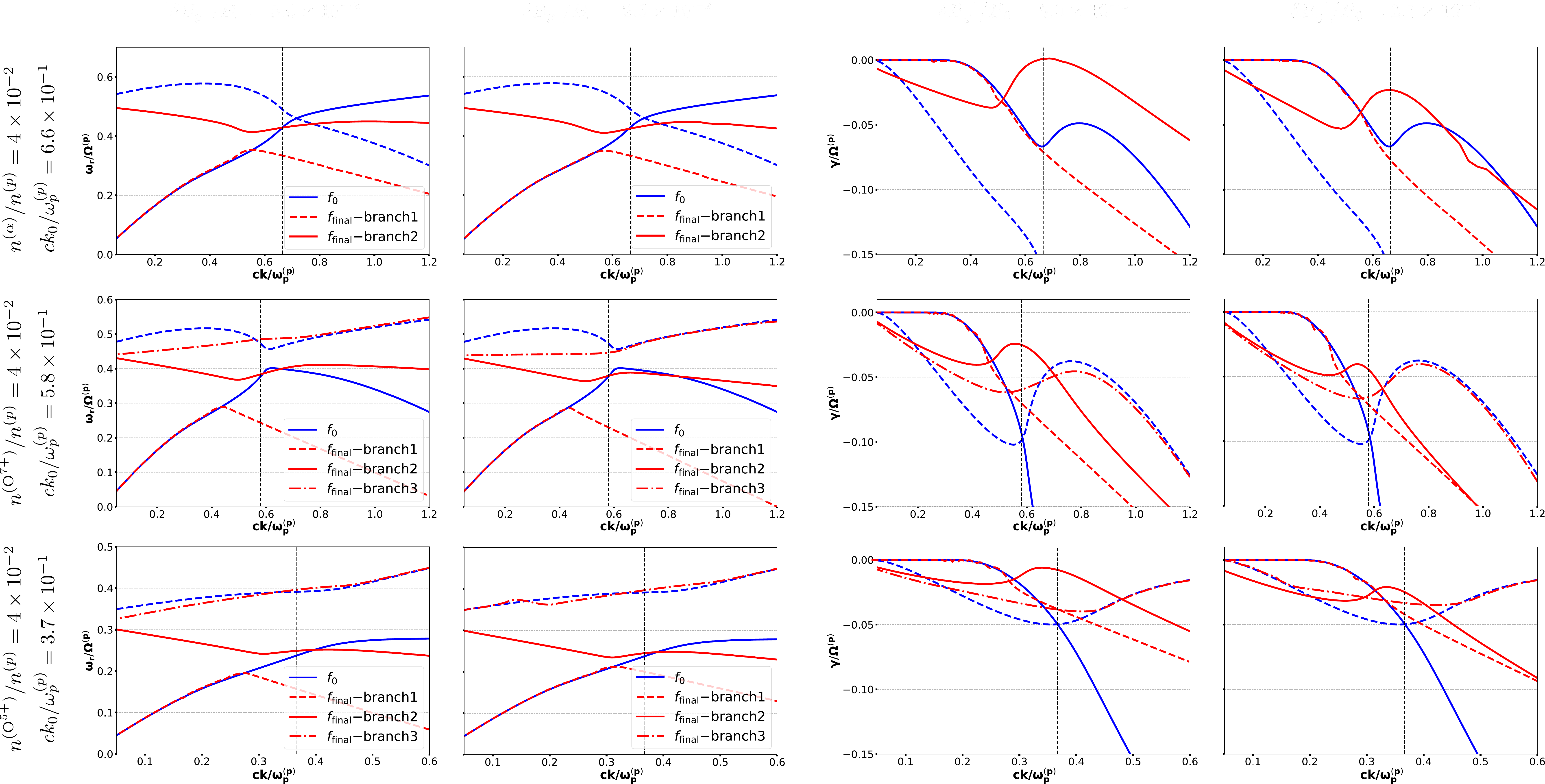}
    \caption{Real (left) and imaginary (right) parts of the dispersion relation of A/IC waves obtained with $\beta^{(s)}=1.6\times 10^{-2}$ for $\alpha$-particles, O${}^{5+}$, and O${}^{7+}$ ions. The results are obtained with the initial Maxwell VDF (blue) and final wave--particle equilibrium VDF (red) with different initial wave amplitudes.  \label{fig:real_w2} }
   
\end{figure*}
Figure~\ref{fig:real_w2} displays the dispersion relations obtained with the initial (blue) and final (red) VDFs with the same ion concentrations as in Figure~\ref{fig:real_w1} but with $\left(v_{\text{th}}^{(s)}/V_{A}^{(p)}\right)^2=0.1$. We simulate the interaction of the test-particles with the lower branch of the Maxwellian plasma dispersion relation, plotted in the figure in continuous blue lines, while the upper branch is plotted in dashed blue lines. We considered  initial wave amplitudes of $|\delta B_y(t=0)|/B_0=2.5\times10^{-2}$ and $|\delta B_y(t=0)|/B_0=5.0\times10^{-2}$.

In Figure~\ref{fig:real_w1}, we use the same wavenumber for all simulations to reduce the parameter space in favor of legibility. In Figure~\ref{fig:real_w2}, the wavenumbers of the forcing wave are different for all species of heavy ions. We chose the wavenumbers $ck_0/\omega_{p}^{(p)}=5.8\times10^{-1}$ for O$^{7+}$ ions and $ck_0/\omega_{p}^{(p)}=3.67\times10^{-1}$ for O$^{5+}$ ions. These are the wavenumbers for which the two branches of the respective Maxwellian electron-proton-Oxygen plasma possess the same damping rate and are associated with the maxima in the heating rate of the test-particles. Since for $\alpha$-particles the upper branch of the dispersion relation is strongly damped, we choose the wavenumber $ck_0/\omega_{p}^{(p)}=6.6\times10^{-1}$ that is related to a local minimum in the damping rate of the lower branch, which is also related to a maximum in the heating rate of this species.

For both of the initial wave amplitudes considered, the dispersion relation of the plasma with $\alpha$-particles splits into two frequency bands. For O$^{7+}$ and O$^{5+}$ ions, the solution splits into three frequency bands. The frequency bands that intersect the lower branch of the initial dispersion relation near $k_0$ are plotted in continuous red lines. For all considered cases, the damping of these solutions is reduced at $ck_0/\omega_{p}^{(p)}$, becoming close to zero for $\alpha$-particles and O$^{5+}$ ions forced with a wave of initial amplitude given by $|\delta B_y(t=0)|/B_0=5.0\times10^{-2}$. In the case of the $\alpha$-particles, the wave becomes slightly unstable at $k_0$.

We now explore the interaction of both Oxygen frequency bands with the test-particle distributions. For $(v_{\text{th}}^{(s)}/V_{A}^{(p)})^2=0.1$, this is omitted for $\alpha$-particles, as already mentioned, since the upper branch is very strongly damped, and thus is not likely to propagate in Maxwellian plasmas. We take O${}^{5+}$ ions as our case study for this thermal speed and consider the wavenumbers for which both frequency bands are equally damped for our initial wave, performing the simulation in electron-proton-O${}^{5+}$ plasmas with $n^{(\text{O}^{5+})}/n^{(p)}=1\times 10^{-2}$ and $n^{(\text{O}^{5+})}/n^{(p)}=5\times 10^{-3}$.
\begin{figure*}[h!]
        \centering
    \includegraphics[width=\textwidth]{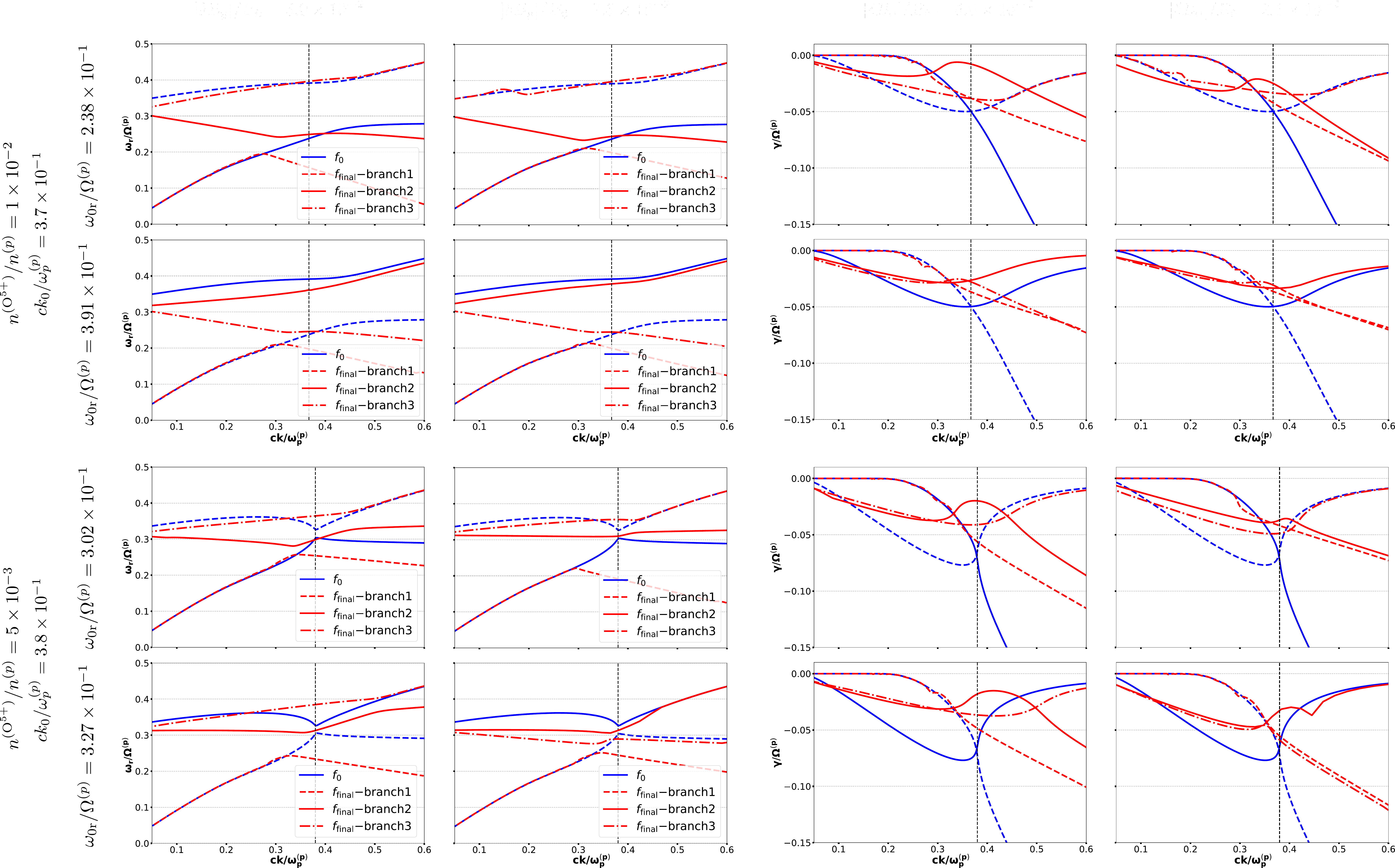}
    \caption{Real (left) and imaginary (right) parts of the dispersion relation of A/IC waves obtained with $\beta^{(s)}=1.6\times 10^{-2}$ for $O^{5+}$ with different densities. We obtain these results with the initial Maxwellian VDF (blue) and final wave--particle equilibrium VDF (red) with different initial wave amplitudes. \label{fig:real_w3} }
    
\end{figure*}
Figure~\ref{fig:real_w3} shows the dispersion relation of a Maxwellian electron-proton-O${}^{5+}$ plasma and for a plasma with the final test-particle distribution of heavy ions for different test-particle concentrations and initial amplitudes of the forcing waves. We use both branches of the dispersion relation in a plasma with Maxwellian distribution in the simulations; we plot the lower branch as continuous and the upper branch as dashed blue lines when the frequency of the forcing wave belongs to the lower branch of the dispersion relation. When the forcing wave belongs to the upper branch, this branch is plotted as continuous, and the lower branch is plotted as dashed blue lines. We plot the dispersion relations obtained with the final VDFs in red, with the one that intersects the initial dispersion relation of the forcing wave at the wavenumber $k=k_0$ as continuous lines. All final VDFs exhibit reduced wave damping at $ck_0/\omega_{p}^{(p)}$ with respect to the damping rate of the Maxwellian plasma dispersion relation, which are collected in Table~\ref{tab:Table1}.

The reduction of the wave damping is more pronounced when $|\delta B_y(t=0)|/B_0=5.0\times10^{-2}$. For this initial wave amplitude, when $n^{(\text{O}^{5+})}/n^{(p)}=1\times 10^{-2}$, the wave damping of the dispersion relation obtained with the final VDF is close to zero when the forcing wave belongs to the lower branch of the Maxwellian plasma dispersion relation, while it approximates $\gamma/\Omega^{(p)}\approx-2\times10^{-2}$ when the forcing wave belongs to the upper branch.  When $n^{(\text{O}^{5+})}/n^{(p)}=5\times 10^{-3}$, the damping rate approximates $\gamma/\Omega^{(p)}\approx-2\times10^{-2}$ when the forcing wave belongs both the lower and the upper branch of the Maxwellian plasma dispersion relation.

\subsection{The effect of drift velocity}\label{sec:drift}
\begin{figure}[h!]
	\centering
	\includegraphics[width=1.0\textwidth]{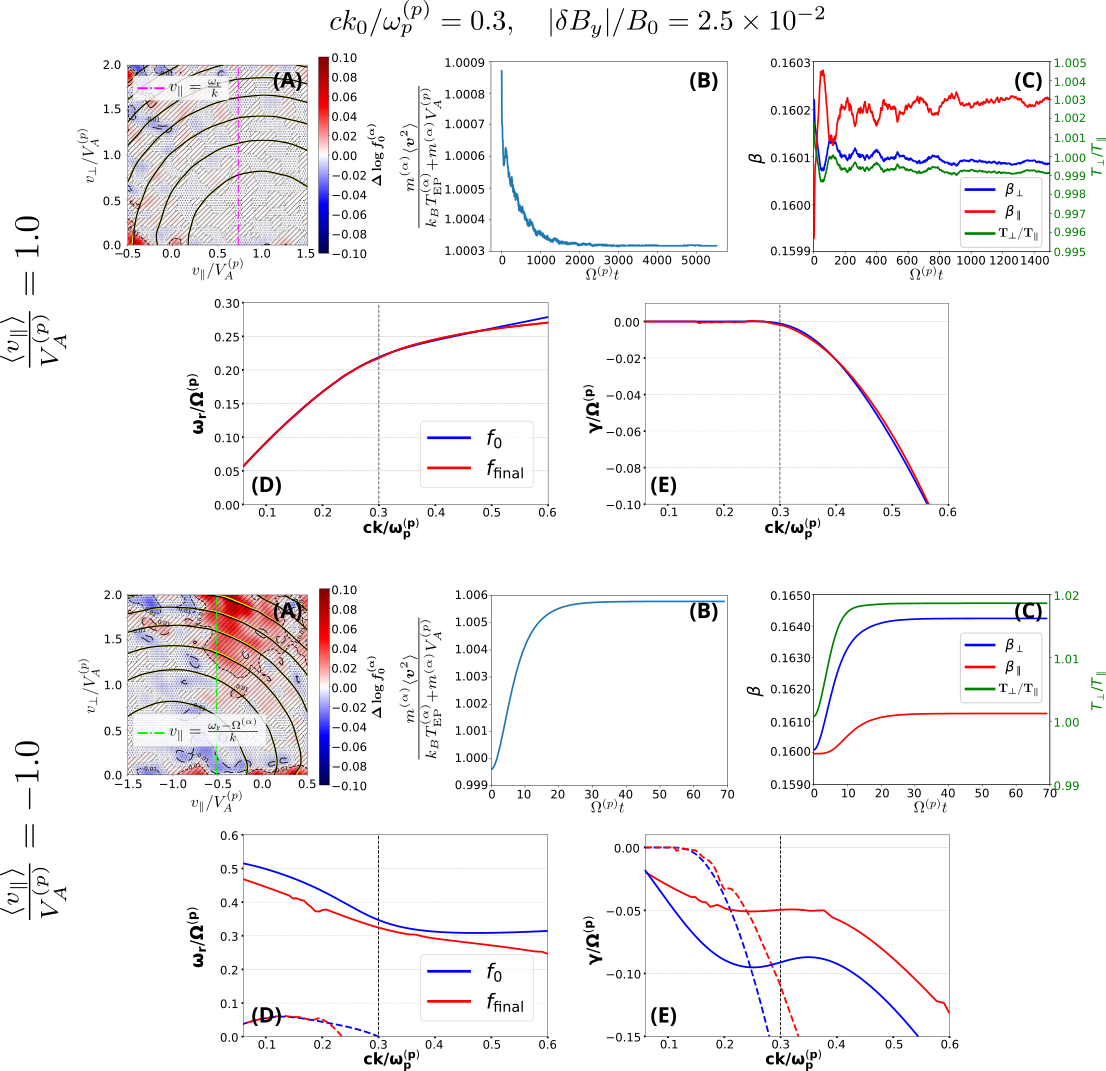}
	\caption{(A) Logarithmic difference between evolved and initial VDF; (B) kinetic energy normalized to thermal plus kinetic energy of the bulk; (C) $\beta$ and anisotropy, (D) real part of dispersion relation of A/IC waves, (E) imaginary part of the dispersion relation; for field-aligned drift velocities of $V_A^{(p)}$ (top) and $-V_A^{(p)}$ (bottom). }\label{fig:drift-f}
\end{figure}

We consider $\alpha$-particle drift velocities of $\pm V_A^{(p)}$ with $\left(v_{\text{th}}^{(\alpha)}/V_{A}^{(p)}\right)^2=1.0$. We display the results of the simulation with $|\delta B_y(t=0)|/B_0=2.5\times10^{-2}$ and $ck_0/\omega_p^{(p)}=0.3$ in Figure~\ref{fig:drift-f}, whose input parameters can be found in Table~\ref{tab:Table1}.\\
The results from the simulation with $\langle v_z\rangle/V_{A}^{(p)}=1.0$ show no significant differences in the VDF due to resonant interaction and decaying trends in both the kinetic energy and temperature anisotropy. Variations in kinetic energy, $\beta_{\perp}$, $\beta_{\parallel}$, and temperature anisotropy, however, are very close to numerical error. The dispersion relation calculated with the evolved VDF at the end of the simulation does not differ significantly from the Maxwellian plasma dispersion relation, with the initial damping rate at the wavenumber $k_0$ being $\gamma/\Omega^{(p)}=-1.14\times10^{-3}$, and $\gamma/\Omega^{(p)}=-1.83\times10^{-3}$ the one obtained with the evolved VDF.\\
For the case that considers $\langle v_z\rangle/V_{A}^{(p)}=-1.0$, the solutions to the dispersion relation representing backward-propagating A/IC show a significant decrease in the damping rate of the resonant branch (continuous blue and red) with respect to the initial, Maxwellian plasma dispersion relation; $\gamma/\Omega^{(p)}=-9.1\times10^{-2}$ for the Maxwellian case, and $\gamma/\Omega^{(p)}=-5.0\times10^{-2}$ for the final VDF. Our results show that $f_{\text{final}}>f_{\text{initial}}$ at large values of $v_{\perp}$ near the resonant field-aligned velocity. The kinetic energy, $\beta_{\perp}$, $\beta_{\parallel}$, and the temperature anisotropy also increase monotonically over the simulation time.  

\subsection{Stability of the final states}
In Figure~\ref{fig:real_w2}, it is shown that the dispersion relation obtained with the evolved distribution for $(v_{\text{th}}^{(\alpha)}/V_{A}^{(p)})^2=0.1$ with $n^{(\alpha)}/n^{(p)}=4\times10^{-2}$ and $|\delta B_y(t=0)|/B_0=5.0\times10^{-2}$ exhibits $\gamma/\Omega^{(p)}\sim0$ at the wavenumber of the monochromatic wave that drives the simulation, $ck_0/\omega_{p}^{(p)}=0.66$. We find  $\gamma(k_0)/\Omega^{(p)}=6.3\times10^{-4}$ as compared to $\gamma(k_0)/\Omega^{(p)}=-6.7\times10^{-2}$ for the Maxwellian plasma dispersion relation. 

We analyze the stability of this system under a new perturbation by a monochromatic wave, which is chosen as the A/IC mode obtained with the evolved VDF of $\alpha$-particles with the previously described parameters. This solution is plotted in continuous red lines in Figure~\ref{fig:real_w2} with $|\delta B_y|/B_0=5.0\times10^{-2}$ for $\alpha$-particles. Since at $ck/\omega_p^{(p)}$ the wave is unstable, we use $ck_0/\omega_{p}^{(p)}=6.4$ for the new forcing wave, which has a damping rate of $\gamma(k_0)/\Omega^{(p)}=-3.1\times10^{-4}$ of the same wavenumber $k_0$.

\begin{figure}
    \centering
    \includegraphics[width=\textwidth]{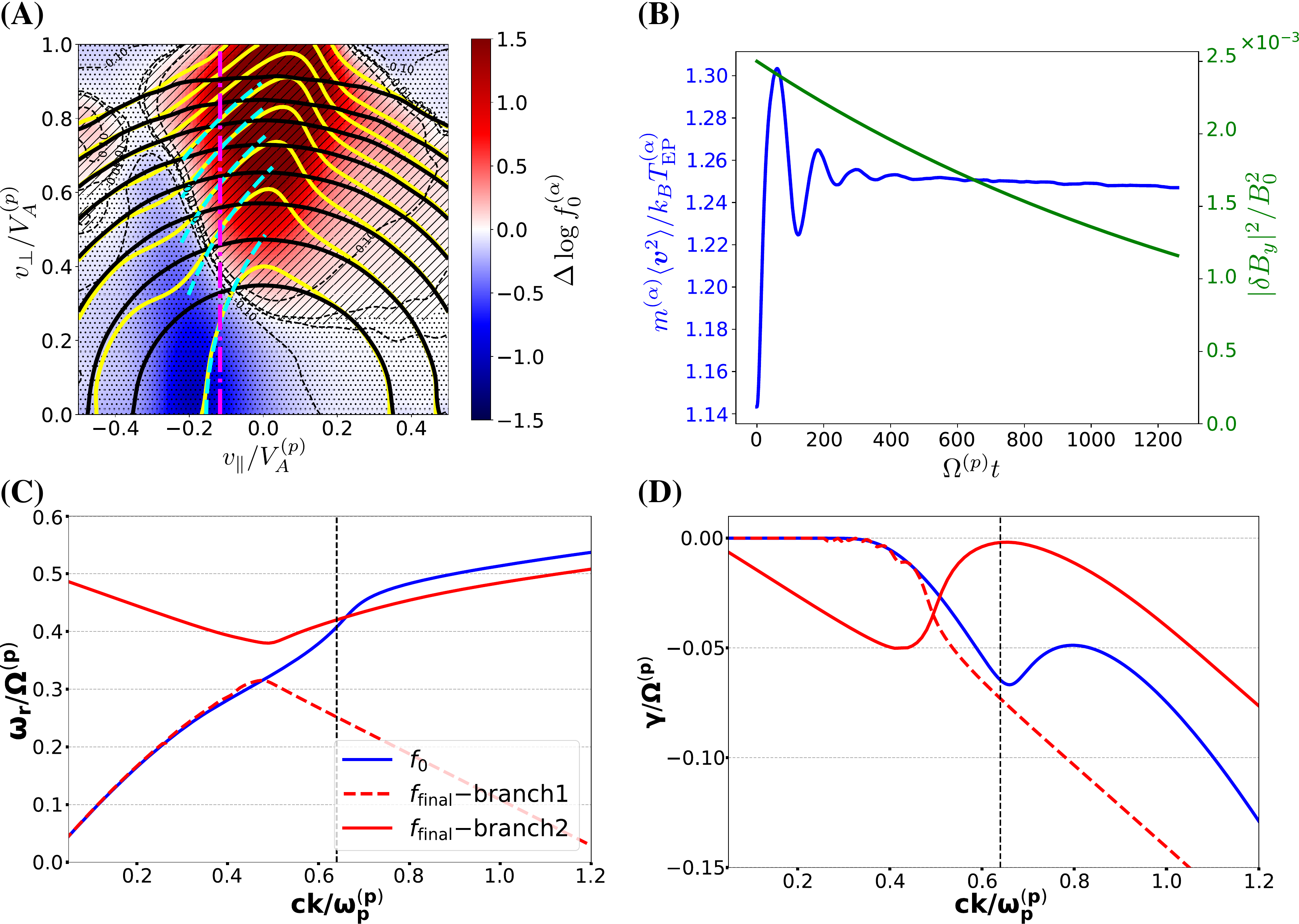}
    \caption{Results from an iteration of the simulation using the evolved VDF dispersion relation for $(v_{\text{th}}^{(\alpha)}/V_{A}^{(p)})^2=0.1$ with $n^{(\alpha)}/n^{(p)}=4\times10^{-2}$, $ck_0/\omega_{p}^{(p)}=0.66$, and $|\delta B_y(t=0)|/B_0=5.0\times10^{-2}$: a) logarithmic difference between evolved VDF (yellow contours) and initial Maxwellian VDF (black contours) near the $v_{\parallel}=(\omega-\Omega^{(\alpha)})/k$ resonance (magenta vertical line), with cyan dashed line segments representing predictions from quasilinear theory; b) mean kinetic energy per particle (blue) and $|\delta B_y|^2/B_0^2$ as functions of time; c) $\omega_{\mathrm r}/\Omega^{(p)}$ as a function of $ck/\omega_{p}^{(p)}$ for a plasma with Maxwellian VDFs (blue) and using the second-iteration evolved VDF for $\alpha$-particles (red); $\gamma/\Omega^{(p)}$ as a function of $ck/\omega_{p}^{(p)}$ using the same color scheme as in c). \label{fig:iteration2}}
   
\end{figure}
Figure~\ref{fig:iteration2} (A) shows the logarithmic difference between the evolved VDF of the second iteration (yellow contours) and the original Maxwellian VDF (black contours) of the $\alpha$-particles. The cyclotron resonance condition is indicated as a magenta dash-dotted line. We superposed predictions from quasilinear theory in cyan. 

Figure~\ref{fig:iteration2} (B) shows the mean kinetic-to-thermal energy ratio of the test-particles (blue) and $|\delta B_y|^2/B_0^2$ (green) as functions of time. The kinetic energy increases from $m^{(\alpha)}\langle v^2\rangle/k_BT_{\mathrm{EP}}^{(\alpha)}\approx1.14$ to $m^{(\alpha)}\langle v^2\rangle/k_BT_{\mathrm{EP}}^{(\alpha)}\approx1.30$  at $\Omega^{(p)}t\approx100$. It decays until it reaches a local minimum at $t\Omega^{(p)}\approx 150$, and oscillates until it approaches a steady value of $m^{(\alpha)}\langle v^2\rangle/k_BT_{\mathrm{EP}}^{(\alpha)}\approx1.24$ for $t\Omega^{(p)}\ge400$. At this point, the wave energy has only decayed by approximately 20\% due to damping. By $\Omega^{(p)}t\approx 1200$, the amplitude of the magnetic field component $|\delta B_y|/B_0$ has decayed to about 50\% of its initial value.  

Panel (C) of Figure~\ref{fig:iteration2} shows the real part of the dispersion relation of A/IC waves calculated using the evolved VDF (red) and the Maxwellian VDF (blue). The solution splits into two frequency bands. At the forcing wavenumber $k_0$, the value of the real frequency of the Maxwellian VDF dispersion relation is $\omega_{r}(k_0)/\Omega^{(p)}=0.41$, while for the upper branch of the evolved VDF dispersion relation, displayed in continuous red lines, it is $\omega_{r}(k_0)/\Omega^{(p)}=0.42$. 

Figure~\ref{fig:iteration2} (D) shows the imaginary part of the dispersion relation of A/IC waves calculated using the evolved VDF (red) and the Maxwellian VDF (blue). At the forcing wavenumber $k_0$, the damping rate of the evolved VDF dispersion relation is $\gamma(k_0)/\Omega^{(p)}=-1.9\times10^{-3}$.

\section{Discussion}
In this article, we introduce a novel method for studying the interactions between A/IC waves and heavy ions in weakly collisional warm plasmas. The physical interpretation of our main findings, their relationship with previous and future observations, and the limitations and further development of our approach are discussed in the following section.

\subsection{Relaxation towards Wave--particle Equilibria}

The framework of quasilinear theory describes the relaxation of a plasma system that is initially unstable to a given kind of plasma wave \citep{Jeong_2020,Yoon2017, Yoon_2024}. In this work, we take a complementary approach. Our results show that, as the damped monochromatic A/IC wave interacts with the heavy ion population, energy is injected locally in velocity space near the $v_{\parallel}$ that satisfies the resonance condition in Equation~\eqref{eq:resonance} with $\ell=1$. This resonant velocity is fully determined by the real part of the wave frequency $\omega_{\mathrm{r}0}$ and the wavenumber $k_0$ of the forcing wave. This energy injection occurs as particles in resonance with the wave are accelerated along directions that are locally tangent to contours of constant energy in the wave frame, leading to modifications of the VDFs of the minor ions that qualitatively resemble the predictions from quasilinear theory as illustrated in Figure~\ref{fig: kin_shell}. As the wave becomes damped, the system reaches a steady state. 

By calculating the self-consistent dispersion relations in plasmas consisting of electrons, protons, and heavy ions, using the evolved VDF obtained from our test-particle simulation for the minor ions and Maxwellian distributions for the other species, we isolate the effect of the nonthermal features of the evolved VDFs in the propagation of A/IC waves. Across all cases examined in this study, Figures~\ref{fig:real_w1}, \ref{fig:real_w2}, and \ref{fig:real_w3} reveal that the damping rates of A/IC waves at wavenumber $k_0$ are consistently lower than those obtained from Maxwellian plasma dispersion relations. In the case with $|\delta B_y|/B_0=2.5\times10^{-2}$ and $\left(v_{\text{th}}^{(s)}/V_{A}^{(p)}\right)^2=1.0$ as well as the case  with $|\delta B_y|/B_0=5.0\times10^{-2}$ and $\left(v_{\text{th}}^{(\alpha)}/V_{A}^{(p)}\right)^2=0.1$, the wave becomes unstable at wavenumber $k_0$, while for all remaining cases the waves remain damped. In these cases, the forcing wave is strongly damped initially so that its amplitude decays significantly before the VDF of the heavy ions relaxes to a transparent state and energy injection ceases. 

This behavior at large wave amplitudes breaks self-consistency in our approach. To obtain a steady state independent of wave damping, we iterate the process using the dispersion relation obtained with the evolved VDF of $\alpha$-particles with $\left(v_{\text{th}}^{(\alpha)}/V_{A}^{(p)}\right)^2=0.1$, which has a damping rate of $\gamma/\Omega^{(p)}\approx -1.9\times10^{-3}$ at $ck_0/\omega_{p}^{(p)}=6.4$. Figure~\ref{fig:iteration2} shows that, in this case, the resulting VDF is consistent with predictions from quasilinear theory near the resonant $v_{\parallel}$, and that the energy injection from the wave to the particles saturates before the wave is significantly damped, suggesting that the ion population has become transparent to the wave. In contrast to the VDF obtained in the first iteration, the resulting VDF does not drive A/IC waves unstable. At wavenumber $k_0$, the damping rate of the wave is of order  $\gamma/\Omega^{(p)}=1\times10^{-3}$, which is sufficiently small to justify treating the wave qualitatively as a normal mode. We therefore identify the VDF depicted in Figure~\ref{fig:iteration2} as a wave--particle equilibrium distribution.

\subsection{The Impact of Relative Drifts and other non-Maxwellian features}

Our work explores the effect of wave--particle interactions in the VDFs of minor ions, which are initially Maxwellian. We show that the particle VDFs evolve towards a wave--particle equilibrium state in which the plasma is transparent for the waves (i.e., $\gamma\to0$) and the VDFs follow a shell-like distribution as the result of quasilinear relaxation.  

In Maxwellian plasmas, the damping of the waves is considerably increased when heavy ions are included compared to the electron-proton case, as shown in Figures~\ref{fig:disp_alpha} and \ref{fig:disp_O5}. The increased damping rate arises because Maxwellian distributions do not correspond to a state of wave--particle equilibrium.

The energy injection to the ion population is directly related to the number of particles that satisfy the cyclotron-resonance condition, which is, in turn, the cause of cyclotron damping in Maxwellian plasmas. Because of this, the introduction of ion drift velocities with respect to the background protons possesses a strong effect on the feasibility of the mechanisms that drive the ions to a wave--particle equilibrium, as observed in our case study of drifting hot $\alpha$-particles $\left(\left(v_{\text{th}}^{(\alpha)}/V_{A}^{(p)}\right)^2=1.0\right)$ in Subsection \ref{sec:drift}. Here, $\omega-\Omega^{(\alpha)}<0$, so a positive drift of $\langle v_{\parallel}\rangle=V_{A}^{(p)}$ moves the core ion distribution far from the resonant velocity. As a result, the effect of the wave on the VDF is negligible, even after long simulation times, due to the small damping rate. 

For $\langle v_{\parallel}\rangle=-V_{A}^{(p)}$, the dispersion splits into two branches, from which the resonant one leads to a similar temporal evolution of the simulation to the nondrifting cases. Since our test-simulation evolves from resonant wave--particle interactions, the effect of velocity drifts is limited to how they modify the dispersion relation of the original VDF. For colder ions with $\left(v_{\text{th}}^{(\alpha)}/V_{A}^{(p)}\right)^2=0.1$, the width of the velocity distributions is smaller, and both negative and positive drifts of the order of $V_{A}^{(p)}$ reduce the number of resonant particles, leading to no significant cyclotron interactions. 

This suggests that the mechanisms discussed in this manuscript are most relevant for backward propagating waves ($\boldsymbol{k}\cdot\langle\boldsymbol{v}\rangle<0$) interacting with hot ion populations in the presence of Alfv\'enic drifts, relevant in the fast and Alfv\'enic slow solar wind.

\subsection{Evolution of the VDF}

Figures~\ref{fig:deltaf}, \ref{fig:heat}, and \ref{fig:iteration2} demonstrate that the process through which a wave--particle equilibrium is reached results in a net gain in kinetic energy of the particles and the development of temperature anisotropies, field-aligned bulk speeds, and heat fluxes, as the VDF evolves through cyclotron resonance. The perpendicular components of odd-order velocity moments, such as bulk velocity and heat flux, vanish, since the resonant wave--particle interactions in the quasilinear approximation preserve the VDF's gyrotropy. Our simulations also confirm this expectation.  

Since the distribution is initially isotropic, the pitch-angle diffusion along constant energy shells in the wave's co-moving frame results in field-perpendicular heating, related to an increase in $T_{\perp}^{(s)}$. Depending on the sign of $v_{\parallel,\text{res}}-\langle v_{\parallel}\rangle$, where $v_{\parallel,\text{res}}=(\omega-\Omega^{(s)})/k_{\parallel}$, the interaction results in an increase ($v_{\parallel,\text{res}}-\langle v_{\parallel}\rangle>0$) or a decrease ($v_{\parallel,\text{res}}-\langle v_{\parallel}\rangle<0$) in $T_{\parallel}^{(s)}$. 

This behavior can be observed in the plots depicting the evolution of $\beta_{\parallel,\perp}$ in Figures~\ref{fig:deltaf} and \ref{fig:drift-f}. In both cases, the temperature anisotropy increases, in consistency with previously reported simulational results \citep{Ofman_2002b,Li_2005,Quijada_2025}. The newly developed non-zero bulk speed of the ions, as well as their temperature anisotropies, are not strong enough to diminish damping significantly or to develop instabilities at $k_0$ if they were simply applied in a bi-Maxwellian model of the dispersion relation. 

Our results show that the changes to the dispersion relation are a product of the local velocity gradients through the non-Maxwellian modifications of the VDF rather than a simple change of the moments of a set of otherwise bi-Maxwellian distributions  \citep{Walters_2023}. 

Under solar wind conditions, $\alpha$-particles with a particular field-aligned velocity, $v_{\parallel}$, may resonate simultaneously with parallel- and antiparallel-propagating A/IC waves \citep{Navarro_2020}. This is likely the cause of their large observed temperature anisotropies and of their differential streaming \citep{Vocks_2002,Kasper_2014}. The co-existence of forward- and backward-propagating A/IC waves may give rise to similar distributions as the dispersive kinetic shells proposed by \cite{Isenberg2004}.

\subsection{Limitations of our Approach}
The relative simplicity of the mechanisms employed in our procedure provides us with a very strong tool to observe the effect of wave--particle interactions on the ion VDFs in a much more direct manner than in self-consistent kinetic simulations. Nevertheless, this comes at the cost of important technical limitations. 

Since the fields do not react to the particles, damping is required for convergence of the simulations; otherwise, there is an infinite reservoir of electromagnetic energy, and the kinetic energy of the test-particles diverges after an initial saturation phase. Therefore, this procedure is only suitable to study damped solutions ($\gamma<0$), whose amplitudes vanish completely after a limited amount of time given by $T=2\pi/|\gamma|$. 

In this framework, the initial amplitude of the wave regulates how much energy is transferred to the test-particles through resonant heating. This amplitude also regulates how quickly the plasma VDFs evolve due to wave--particle interactions within the framework of quasilinear theory through the scaling of the right-hand side of Equation~\eqref{eq:QL}, and therefore has a significant effect on the achievement of wave--particle equilibrium states. As seen in Figures~\ref{fig:real_w1}, \ref{fig:real_w2}, and \ref{fig:real_w3}, a smaller initial amplitude produces damped final states in which damping is reduced with respect to the Maxwellian dispersion relation. 

When the initial wave amplitude is large, it may not strictly obey the linear assumption $|\delta B|\ll B_0$. Here, the wave heats the particles through paths with steeper slopes in $v_{\parallel}-v_{\perp}$ space than those of quasilinear diffusion, which can be observed in Figure~\ref{fig: kin_shell}. The resulting nonthermal velocity-space gradients in the VDF in the vicinity of the resonant velocity can lead to kinetic-scale instabilities \citep{Walters_2023,Klein_2025}, as seen in the large-amplitude cases depicted in Figures~\ref{fig:real_w1}, \ref{fig:real_w2}, and \ref{fig:real_w3}. 

Observed plasma waves are not monochromatic but rather occur over a range of wavenumbers and in multiple frequency bands \citep{Saikin_2015}. Thus, we expect our results to be observable as a development of a local anisotropy in the heavy ion VDF over a range of $v_\parallel$ values, suggesting the development of wave--particle equilibria over a range of wavenumbers. Similar kinetic steady states also occur in particle-in-cell (PIC) simulations of Landau resonance of Langmuir waves with initially Maxwellian VDFs \citep{Carril_2023}. 

The method introduced in this study is not limited to monochromatic A/IC waves. However, the presence of fully nonlinear electromagnetic turbulence is not applicable within this model, as it is not suitable for reproducing wave-wave interactions. Therefore, a fully nonlinear study of steady-state wave particle equilibria must be developed using kinetic plasma simulations, as has been done by \cite{martinovic2026}, in direct comparison with ALPS. This, however, lies beyond the scope of this article.

\section{Conclusions}

The study of resonant interactions between electromagnetic waves and minor ions is a matter of great interest to our understanding of particle dynamics and heating in the solar wind. Their gyrofrequency, which is generally lower than the proton gyrofrequency, allows heavy ions to interact with low-frequency A/IC waves \citep{Hollweg_2002,Moya_2014, Navarro_2020}, like those that are observed in the solar wind between 0.045 and 1 au \citep{Jian_2010, Jian_2014,  Bowen_2020, Niranjana_2024}.

Minor ions are often treated as test particles when discussing the acceleration and heating of these particles in the solar wind \citep{Chen2002,Kasper_2014,Pugliese2022}. Indeed, test-particle simulations are a useful tool for studying the interaction of waves with these ions and their effects on the propagation and dispersion relation of kinetic plasma waves.  

With this work, we are able to reproduce the development of ion temperature anisotropy, a feature that is commonly observed in the solar wind and associated with wave activity \citep{Feldman_1973,Marsch1982c}. Our results suggest that the VDFs of minor ions that undergo cyclotron interactions with parallel-propagating electromagnetic waves are likely to develop skewness and an associated heat flux. This is consistent with the recent observation that A/IC waves can drive a small field-aligned proton heat flux \citep{Niranjana_2025}. Similar features are also observed in the VDFs of $\alpha$-particles in the solar wind \citep{DeMarco2023,Perrone_2024}. We also show that small field-aligned relative drift velocities are more efficient at producing states with $T_{\perp}^{(\alpha)}/T^{(\alpha)}_{\parallel}>1$ due to interaction with forward-propagating waves, in agreement with recent observations \citep{Jagarlamudi_2025}.

This simple approach provides insight into wave–driven kinetic processes that naturally occur in weakly collisional plasmas, their subsequent modification of ion VDFs, and how this modification allows kinetic-scale electromagnetic wave propagation that would otherwise be heavily damped.
This insight contributes to our understanding of wave–particle interactions and wave propagation in space and astrophysical plasma environments. Other future efforts can also consider a more realistic setup that incorporates both $\alpha$-particles and Oxygen ions, which could be used in future studies to quantify the relative significance of wave–-particle interactions involving both heavy ion species simultaneously.

\begin{nolinenumbers}
	\section*{Acknowledgments}
		N.V.S. gratefully acknowledges the support provided by the National Agency for Research and Development (ANID) of Chile through the National Doctoral Scholarship 21220616. This scholarship partially funded an internship at the Mullard Space Science Laboratory (MSSL), University College London (UCL). N.V.S. extends sincere gratitude to the staff of MSSL for their hospitality and support during the internship, which greatly contributed to the progress of this work. D.V. is supported by STFC Consolidated Grant ST/W001004/1. We also thank the support of ANID, Chile, through FONDECyT grants No. 1240281 (P.S.M.) and 1251712 (R.A.L). K.G.K. was supported by NASA grant 80NSSC24K0724.  We also thank Dr. Abiam Tamburrini (Universidad de Chile), Dr. Jesse Coburn (MSSL-UCL), and Mr. Diego Rodríguez Cid (Pontificia Universidad Católica de Chile) for fruitful discussions.  The ALPS project received support from UCL’s Advanced Research Computing Centre through the Open Source Software Sustainability Funding scheme. This study benefited from support by the International Space Science Institute (ISSI) in Bern, through ISSI International Team project 24-612 (``Excitation and dissipation of kinetic-scale fluctuations in space plasmas'').
\end{nolinenumbers}

\bibliography{bibliography}{}
\bibliographystyle{aasjournal}

\end{document}